\documentclass[reprint,
 amsmath,amssymb,
 aps,
 pre,
]{revtex4-2}

\usepackage{graphicx}
\usepackage{dcolumn}
\usepackage{bm}
\usepackage{hyperref}
\usepackage{xcolor}
\usepackage{cancel}
\usepackage{lipsum}
\usepackage{stmaryrd} 

\begin{document}

\title{Nonreciprocal random networks and their percolation properties}

\author{Chanania Steinbock}
\email{chanania.steinbock@jhu.edu}
\affiliation{Department of Physics and Astronomy, Johns Hopkins University, Baltimore, MD 21218,
USA}

\date{\today}

\begin{abstract}
We study the effects of nonreciprocity and network structure on percolation. To this end, we investigate nonreciprocal random networks -- directed networks for which the probability of a link occurring from node $i$ to node $j$ differs from the probability of the reverse link occurring from node $j$ to node $i$. We analytically determine the degree and percolation properties of such networks with exactly two types of link probability, demonstrating that whether the networks are structured such that the nodes are not statistically indistinguishable has profound effects on these measures, both quantitively and in how such networks need to be approached. In particular, we develop a technique for solving the percolation problem which can be applied to both structured and unstructured networks. The method entails writing self-consistent integral and differential equations for the probability that each node will belong to the network's giant component. Exact solutions to these equations are obtained and simulations which confirm our analytic predictions are presented.
\end{abstract}

\maketitle


\section{\label{sec:Intro}Introduction}

Random networks have become an indispensable component of network science since their introduction by Paul Erd\H{o}s, Alfr\'{e}d R\'{e}nyi and Edgar Gilbert in 1959 \cite{Erdos1959, Gilbert1959}. In particular, their tendency to exhibit percolation transitions, in which a macroscopic fraction of the network forms an emergent structure, has been a rich prototype for understanding complex systems, with applications to social and citation networks, ecosystems and biological networks, trade and transportation networks, the internet and power networks, and much more \cite{WassermanBook, NewmanBook, EstradaBook, BarabasiBook}.

Though less studied, directed networks, in which each link has a source node and a target node, are an important variant for studying systems which lack symmetric relations between nodes. For instance an individual may follow someone else's social media account while not being followed by them in turn \cite{Gupte2011, Samanta2021}, or a website may link to another site while not having the relationship be reciprocated \cite{Albert1999, Tadic2001}. In some instances, the relationship cannot be reciprocated, even in principle. For instance, in citation networks, a later paper can cite earlier ones but never the reverse \cite{Redner1998, Newman2001Citation, Steinbock2019DSPL, Steinbock2019Deg}.

A key measure to describe the ``degree of directedness'' of a random network is the reciprocity \cite{WassermanBook, NewmanBook, EstradaBook, Newman2002Email, Serrano2003, Garlaschelli2004}. This measures the fraction of node pairs which reciprocate links between themselves. When the reciprocity is large such that most connected nodes reciprocate links, little is lost by approximating the network as undirected. In contrast, when the reciprocity is low, the directed character cannot be neglected. 

In \cite{Garlaschelli2004}, it was observed that real world networks from similar contexts exhibit similar types of reciprocity. For instance, social, trade and biological networks tend to exhibit positive reciprocity indicating that they show greater reciprocity than a random network would, while financial networks and ecosystems tend to exhibit negative reciprocity, indicating that they are less likely than a  corresponding random network to exhibit reciprocal relationships. Accordingly, various models have been proposed with the explicit goal of capturing this nonreciprocity \cite{Holland1981, Park2004, Meyers2006}.

Indeed, over the years, many random network models of varying degrees of generality have been developed. Let us just mention a few which will bear on the subject of this paper. Arguably, the most general model conceived that retains probabilistic independence between links is the ensemble in which each link in the network, say from node $i$ to node $j$, exists independently with some given probability $P_{i\rightarrow j}$ \cite{Bollobas2007, McCulloh2007}. While \cite{Bollobas2007}  defines this model, $G(N,\{P_{i\rightarrow j}\})$, for undirected networks on $N$ nodes, the generalisation to the directed case is straight-forward, as demonstrated in \cite{McCulloh2007}. In truth little is known about this model and as pointed out in \cite{Bollobas2007}, ``it seems difficult to obtain substantial asymptotic results \ldots without further restrictions; the model is too general''. Of course, the simplest restriction that can be applied to this model is the one in which $P_{i\rightarrow j} = p$ for all $i$ and $j$ and the undirected variant of this on $N$ nodes, often denoted $G(N,p)$, is just the classic random network first studied in \cite{Gilbert1959}.

More interesting restrictions which have been profitable can be categorised as ``kernel networks'' \cite{Bollobas2007, Cao2020} or ``multi-type networks'' (also known as ``stochastic blockmodels'') \cite{Holland1983, Bradde2009, Bloznelis2012, Allard2015}. In a kernel network, the link probability for each pair of nodes $i$ and $j$ is generated by some function $P_{i\rightarrow j}=\kappa(i,j)$ whereas in a multi-type network, each node is assigned a type $\nu(i)$ and the probability of connecting nodes of each type $\nu$ and nodes of each type $\nu'$ is specified. Such models are closely related as the kernel $\kappa(i,j)$ is often nothing more than a function of the types $\nu(i)$. Note that in the limit where the number of types equals the network size that the fully general model is retrieved.

These models are sufficiently general to include many specific random network models while also being sufficiently specific to enable concrete results to be asserted. In particular, they are sufficiently general to be able to describe structured networks in which nodes throughout the network are not statistically indistinguishable. At the same time, due to their great generality, there are severe limits to how much they can assert. As pointed out in \cite{Bradde2009}, ``the general case of the structured networks \ldots has to be solved on a case-by-case basis''.

In this paper, we will focus on a specific subclass of these general networks, characterised by an emphasis on nonreciprocity. This subclass is interesting because it is both sufficiently specific for its percolation properties to be exactly solvable while also being sufficiently general to allow unstructured, partially structured and fully structured networks to be considered simultaneously. This makes it an ideal candidate for investigating both the effect of network structure and the effect of nonreciprocity on percolation. 
Similar models of nonreciprocal directed networks on various kinds of lattice have previously been fruitfully studied, where they have variously been called \textit{resistor-diode networks} \cite{Redner1981, Redner1982, Inui1999, Janssen2000}, \textit{biased directed percolation models} \cite{Zhou2012, deNoronha2018} or simply \textit{mixed graphs} \cite{Verbavatz2021}. 

Over the last few decades, a number of techniques have been developed to study percolation in random networks \cite{Li2021}. Important examples include the generating function method \cite{Newman2001, Dorogovtsev2001, Schwartz2002, Boguna2005, Meyers2006}, the message passing method \cite{Shiraki2010, Karrer2014, Newman2014, Allard2015, Qian2024}, reduction to the Potts model \cite{Stephen1976, Wu1978, Dorogovtsev2004, Lee2004, Bradde2009} and the use of kinetic rate equations \cite{KrapivskyBook, Krapivsky2001, Dorogovtsev2001Growing, Kim2002, Ben-Naim2005}. Each of these methods has its advantages and disadvantages. The generating function method is simple to apply and can produce analytic predictions but has limited applicability to structured networks. In contrast, the message passing method is capable of handling much more general networks, including real networks, but its results tend to be more numerical than analytical. Reduction to the Potts model often entails exchanging one difficult problem for another while kinetic rate equations are only appropriate for models which can be recast as dynamic growth processes. 

In this paper, we study percolation differently. Since we are interested in treating both structured and unstructured networks analytically, the generating function method and message passing methods are largely inapplicable. Instead, similar to the message passing method, we write self-consistent integral and differential equations for the probability that a given node $i$ can reach either another given node $j$ or the network's giant component (assuming one exists). For the model we consider in this paper, these equations are solvable exactly.

An outline of the paper is as follows. In Sec.~\ref{sec:Nonreciprocal Random Networks}, we define the nonreciprocal random network model and describe its main structural features. Then, in Sec.~\ref{sec:Degree Distribution}, we calculate its most fundamental degree features and show how these behave very differently when the networks are structured. Its percolation properties are investigated in Sec.~\ref{sec:Percolation} and numerical simulations are presented in Sec.~\ref{sec:Simulations}. A discussion of our results is presented in Sec.~\ref{sec:Discussion} and the details of various non-central calculations are relegated to Appendices~\ref{sec:Reciprocity} and \ref{sec:GF Method}.

\section{\label{sec:Nonreciprocal Random Networks}Nonreciprocal Random Networks}

Consider a directed simple network which consists of $N$ nodes. For each pair of nodes, $i$ and $j$, let $P_{i\rightarrow j}$ denote the probability that node $i$ is connected to node $j$. We define a \textit{nonreciprocal random network} to be any network for which $P_{i\rightarrow j} \ne P_{j\rightarrow i}, \forall i\ne j$.

\begin{figure}
    \centering
    \includegraphics[width=0.8\linewidth]{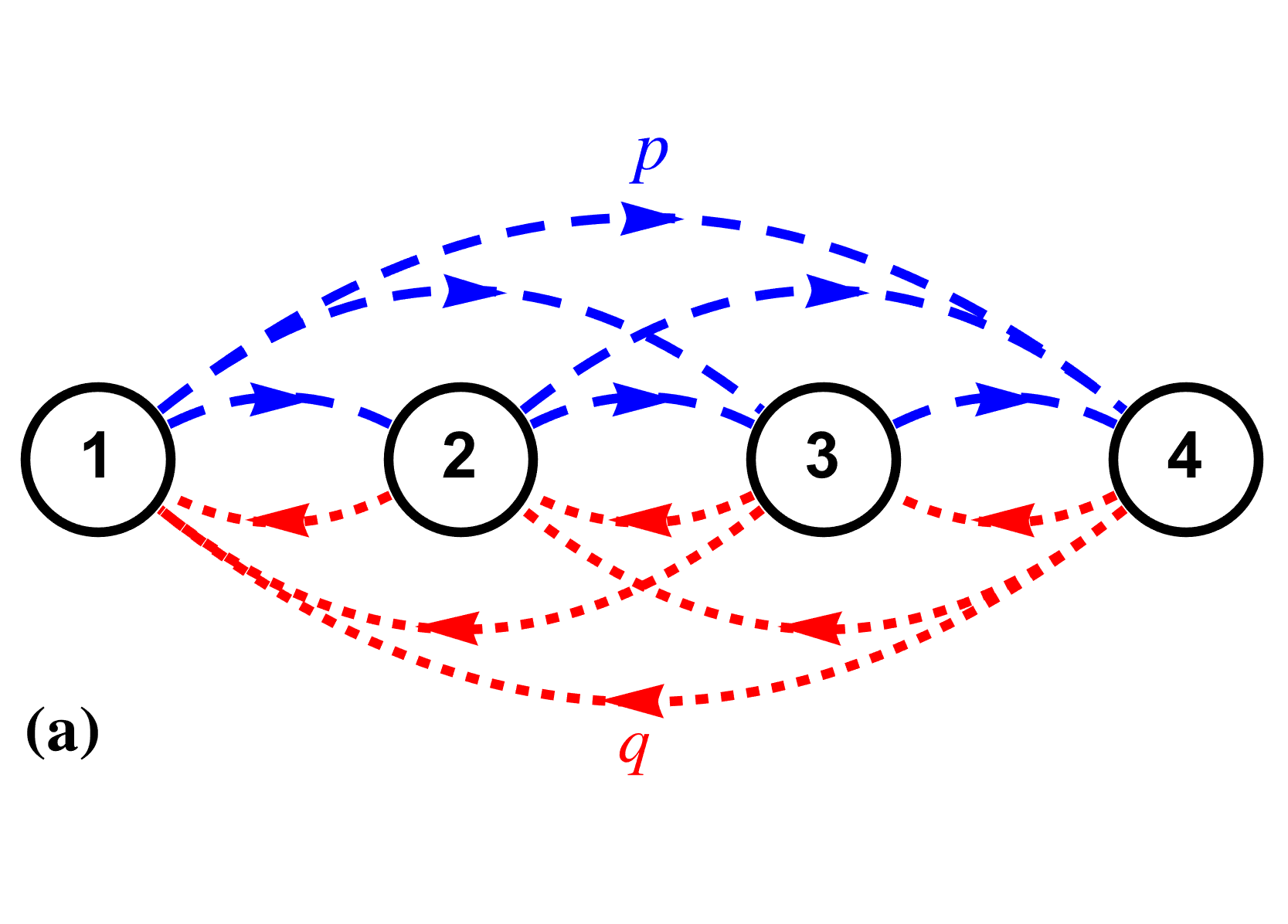}
    
    \vspace{0.1 cm}
    
    \includegraphics[width=0.5\linewidth]{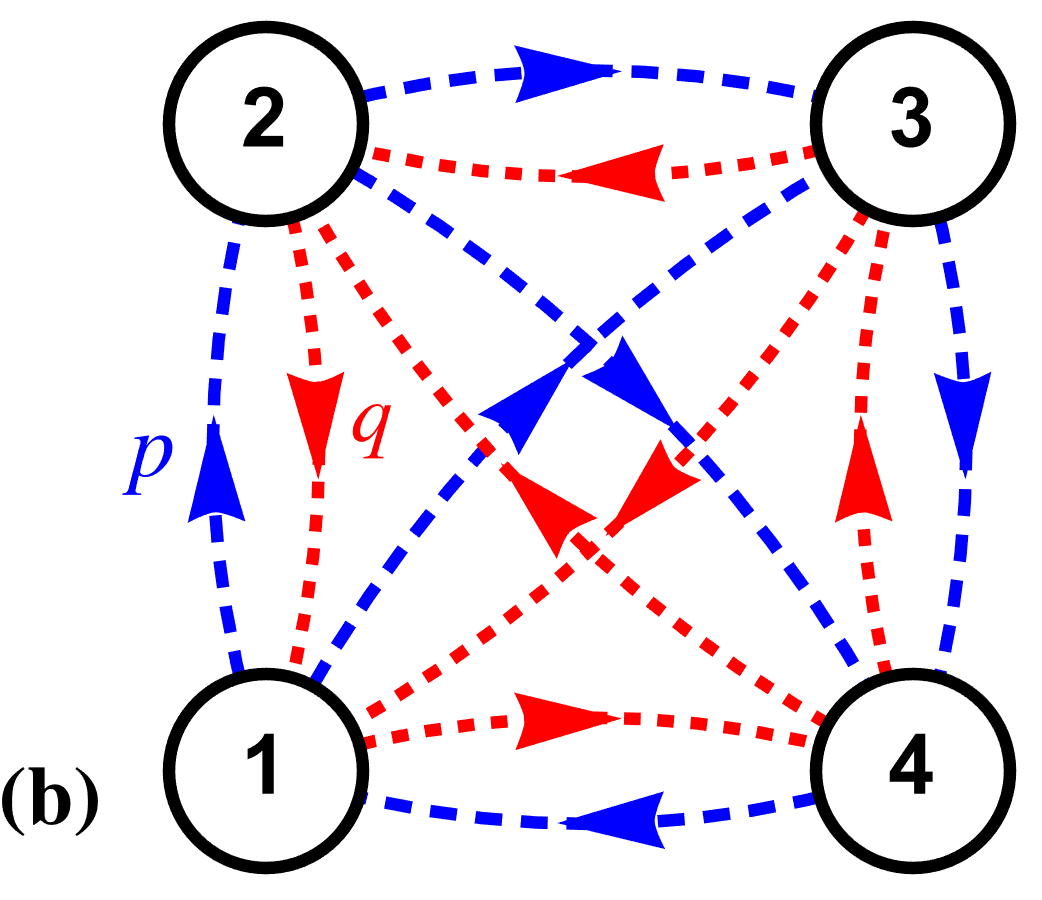}
    
    \caption{\textbf{(a)} A transitive structured nonreciprocal random network on 4 nodes. \textbf{(b)} An example of an intransitive unstructured nonreciprocal random network on 4 nodes. In each case, the blue links (long dashes) exist with probability $p$ while the red links (short dashes) exist with probability $q$.}
    \label{fig:Example networks}
\end{figure}

The matrix $P_{i\rightarrow j}$ is known as the \textit{network probability matrix} \cite{McCulloh2007}. For convenience, we will only treat networks without self-loops such that the diagonal elements can be taken to be zero, i.e. $P_{i\rightarrow i} = 0, \forall i$. In the simplest case, the network probability matrix consists of exactly two types of probability, say $p$ and $q$, such that for each pair of nodes $i$ and $j$, $P_{i\rightarrow j} = p$ and $P_{j\rightarrow i} = q$. In this paper, we will limit our attention to such cases.

We will call a network \textit{structured} if the network probability matrix depends on $i$ or $j$. As an example, if we number the nodes $1,2,...,N$ and have
\begin{equation}
    P_{i \rightarrow j} = 
    \begin{cases}
        p & i<j\\
        q & i>j
    \end{cases} \quad,
    \label{eq:P_ij transitive}
\end{equation}
then the resulting network is structured. In this particular case, the network probability matrix has the highly ordered form
\begin{equation}
    P_{i\rightarrow j} = 
    \left(\begin{array}{ccccc}
    0 & p & p & p & \ldots\\
    q & 0 & p & p & \ldots\\
    q & q & 0 & p & \ldots\\
    q & q & q & 0 &       \\
    \vdots & \vdots & \vdots &  & \ddots
\end{array}\right) \,.
\end{equation}

A nonreciprocal random network can also be completely unstructured. For instance, for each pair of nodes $i$ and $j$ with $i<j$, randomly choose between $P_{i\rightarrow j} = p$ and $P_{i\rightarrow j} = q$. This determines the upper triangular part of the network probability matrix which in turn completely determines the lower triangular part.

The structured network defined by Eq.~(\ref{eq:P_ij transitive}) has the additional property of being \textit{transitive} in the following sense. If 
$P_{i\rightarrow j} = p$ and $P_{j\rightarrow k} = p$, then 
$P_{i\rightarrow k} = p$ (and similarly if we replace $p$ with $q$). Accordingly, the nodes in the network can be ordered by their probability of reaching ``higher'' or ``lower'' nodes. In contrast the completely unstructured nonreciprocal random network defined in the previous paragraph is completely intransitive. Examples of these two kinds of network are schematically shown in Fig.~\ref{fig:Example networks}. In general, the degree of transitivity of a network can vary between completely intransitive and completely transitive. For instance, instead of randomly selecting evenly between $p$ and $q$ in the unstructured example above, one can select $p$ with some probability $(1+t)/2$ and $q$ with probability $(1-t)/2$,  ($t\in[-1,1]$). This will create a bias for or against $p$ in the upper-triangular part of the network probability matrix thus creating a tendency towards transitivity. When $|t| = 1$, complete transitivity is obtained while $t = 0$ corresponds to complete intransitivity. A natural notation for these networks, consistent with the notations mentioned in the introduction, might be $G(N,\{p,q\};t)$. Of course, there are many other possible structures one can envision. Determining a complete set of parameters, like the transitivity $t$, for characterising structure is beyond the scope of this paper. Here, we will limit our focus to the two extreme cases of $G(N,\{p,q\};t=0)$ and $G(N,\{p,q\};t=1)$, and not consider networks with intermediate transitivity. 

A note on terminology -- in previous works (for instance \cite{WassermanBook}) the term ``transitive'' has been used to describe a stronger condition; three nodes $i$, $j$ and $k$ are transitive if $i$ is connected to $j$ and $j$ is connected to $k$ implies $i$ is connected to $k$. In contrast, our usage of the term is probabilistic,  $i$, $j$ and $k$ are transitive if $P_{i\rightarrow j} = p$ and $P_{j\rightarrow k} = p$ implies $P_{i\rightarrow k} = p$. This latter notion of transitivity is more useful for our purposes as it determines a (partial) ordering of the nodes and thus illuminates the network structure. In contrast, since $i\rightarrow j$ and $j\rightarrow i$ are not mutually exclusive, the former definition doesn't determine ordering but rather equivalence. Indeed, the former definition of transitivity is now-a-days contained in the more general and useful concept of \textit{clustering} in networks \cite{NewmanBook, BarabasiBook}. Accordingly, in this paper, ``transitivity'' will exclusively be used to refer to the probabilistic sense defined in the previous paragraph.

Two nodes $i$ and $j$ are considered reciprocal in a network if there exists both a link from $i$ to $j$ and a link from $j$ to $i$ \cite{Garlaschelli2004}. The extent to which a given network with adjacency matrix $A$ is reciprocal can be characterised by the reciprocity \cite{Garlaschelli2004}
\begin{equation}
    \rho[A]  = 
    \frac{\sum_{i\ne j}(A_{ij}-\bar{A})(A_{ji}-\bar{A})}
    {\sum_{i\ne j}\left( A_{ij}-\bar{A}\right)^2}
    = \frac{r - \bar{A}}{1 - \bar{A}}\,,
\end{equation}
where $\bar{A} = \sum_{i\ne j}A_{ij}/N(N-1)$ denotes the average of $A$ and $r = L^\leftrightarrow/L$ denotes the ratio of the number of reciprocal links in the network, $L^\leftrightarrow = \sum_{i\ne j}A_{ij}A_{ji}$, to the total number of (non-self) links in the network, $L = \sum_{i\ne j}A_{ij}$. For a random network, the adjacency matrix $A$ is a random variable and thus $\rho$ too will be a random variable.

In Appendix~\ref{sec:Reciprocity}, we calculate the mean reciprocity $\left<\rho\right>$ for \textit{any} nonreciprocal random network on $N$ nodes with exactly two probabilities $p$ and $q$. For even moderately large $N$, the result is remarkably found to not depend on the size of the network
\begin{equation}
    \left<\rho\right>\approx
    -\frac{(p-q)^{2}}{(p+q)(2-p-q)}\,.
    \label{eq:mean reciprocity}
\end{equation}
Since $\left<\rho\right> < 0$, we find that nonreciprocal random networks are ``anti-reciprocal'', i.e. they are less likely to have reciprocal pairs of nodes than a corresponding random network (obtained for example by randomly redistributing the links throughout the entire network). This makes sense. If we randomly redistribute the links throughout the network, the probability of there existing a link between any pair of nodes will be $(p+q)/2$ thus the probability of two nodes being reciprocal to each other is $[(p+q)/2]^2$. In contrast, in the nonreciprocal random network, the probability of any node pair being reciprocal is $pq$. Since the inequality of arithmetic and geometric means ensures that $(p+q)/2 \ge \sqrt{pq}$, it makes sense that nonreciprocal random networks are anti-reciprocal relative to a corresponding random network.

In the following sections, we will calculate the degree distributions and percolation properties of the transitive and intransitive nonreciprocal random networks defined above. Comparison between them will yield valuable insight into the effect of network structure on network characteristics.

\section{\label{sec:Degree Distribution}Degree Distributions}

\subsection{\label{sec:Intransitive Degree Distribution}Intransitive Networks}

For the unstructured intransitive network, each node is statistically identical and thus the network degree distribution is characterised by the two distributions 
\mbox{$P(K_\mathrm{out} = k)$} and $P(K_\mathrm{in} = k)$ corresponding to the out-degree distribution and in-degree distribution of a randomly selected node.

Consider some node $i$. The probability that this node will be connected to some other node $j$ is equal to the weighted probability of $p$ and $q$ which for the completely unstructured case is just $(p+q)/2$. Accordingly, the out-degree of node $i$ is just binomially distributed over the remaining $N-1$ nodes with probability $(p+q)/2$, i.e.
\begin{multline}
    P(K_\mathrm{out} = k) = \\
    \binom{N-1}{k}
    \left(\frac{p+q}{2}\right)^k
    \left(1 - \frac{p+q}{2}\right)^{(N-1)-k} \,.
    \label{eq:P k_out intransitive}
\end{multline}
The exact same reasoning applies to the in-degree distribution and thus we have
\begin{equation}
    P(K_\mathrm{in} = k) = P(K_\mathrm{out} = k) \,.
    \label{eq:P k_out = P k_in intransitive}
\end{equation}

In the continuation, we will be interested in taking the large $N$ limit. To this end, we define the network parameters $\lambda_p$ and $\lambda_q$ via the relationships
\begin{align}
    p = \frac{\lambda_p}{N-1} \,, \label{eq:lambda_p def}\\
    q = \frac{\lambda_q}{N-1} \,. \label{eq:lambda_q def}
\end{align}
$\lambda_p$ can be thought of as the average number of nodes a randomly chosen node will connect to due to forming links with probability $p$ (and similarly for $\lambda_q$). In the large $N$ limit holding $\lambda_p$ and $\lambda_q$ constant, the degree distributions $P(K_\mathrm{out} = k)$ and $P(K_\mathrm{in} = k)$ become Poisson distributions
\begin{multline}
    P\left(K_{\mathrm{out}}=k\right) = 
    P\left(K_{\mathrm{in}}=k\right) = \\
    \frac{1}{k!}\left(\frac{\lambda_{p}+\lambda_{q}}{2}\right)^{k}e{}^{-\frac{\lambda_{p}+\lambda_{q}}{2}} \,,
    \label{eq:P k out in large N}
\end{multline}
and thus
\begin{multline}
    \left<K_\mathrm{out}\right> = 
    \left<K_\mathrm{in}\right> = \\
    = \mathrm{Var}[K_\mathrm{out}] = 
    \mathrm{Var}[K_\mathrm{in}] = 
    \frac{\lambda_p+\lambda_q}{2} \,.
    \label{eq:degree statistics intransitive}
\end{multline}

So far, regarding degree distributions, the unstructured nonreciprocal random network is indistinguishable from an ordinary directed random network with link probability $(p+q)/2$. Differences emerge however when we consider correlations. For instance, consider the joint distribution \mbox{$P(K_\mathrm{out} = k_\mathrm{out}, K_\mathrm{in} = k_\mathrm{in})$} that a randomly chosen node will have out-degree $k_\mathrm{out}$ and in-degree $k_\mathrm{in}$. For an ordinary random directed network, the in-degree and out-degree will be independent and thus the joint distribution is just the product of the two marginal distributions. In contrast, for the  nonreciprocal random network, this independence fails. To gain some intuition for why this must be so, suppose that $p$ is very large ($1-p\ll1$) and $q$ is very small ($q\ll1$) and suppose we learn the out-degree of node $i$ is $K_\mathrm{out}$. From this, we can infer with high probability that node $i$ attempted to link to roughly $K_\mathrm{out}$ nodes with probability $p$ and $N-1-K_\mathrm{out}$ nodes with probability $q$. Accordingly, those $K_\mathrm{out}$ nodes attempted to link to node $i$ with probability $q$ and those $N-1-K_\mathrm{out}$ nodes attempted to link to node $i$ with probability $p$. Since $p$ is large and $q$ is small, we can infer that $K_\mathrm{in} \approx N-1-K_\mathrm{out}$. We thus find that learning the value of $K_\mathrm{out}$ provides information about the value of $K_\mathrm{in}$ and thus these quantities are not independent.

To calculate the joint degree distribution, consider a node $i$ and the $N-1$ other nodes in the network. Each other node can be classified as either being of ``type-$p$'' or ``type-$q$'' relative to node $i$ depending on whether the probability of being linked to is $p$ or $q$. Now suppose that node $i$ is related to $N_p$ type-$p$ nodes and $N_q$ type-$q$ nodes. Then we can define the joint conditional probability $P(k_\mathrm{out},k_\mathrm{in}|N_p,N_q)$ that a randomly selected node will have out-degree $k_\mathrm{out}$ and in-degree $k_\mathrm{in}$ given that it is in relation to $N_p$ type-$p$ nodes and $N_q$ type-$q$ nodes. The advantage of defining this quantity is that $P(k_\mathrm{out},k_\mathrm{in}|N_p,N_q)$ factorises, i.e.
\begin{equation}
    P(k_\mathrm{out},k_\mathrm{in}|N_p,N_q) = P(k_\mathrm{out}|N_p,N_q)P(k_\mathrm{in}|N_p,N_q) \,.
\end{equation}
This is because knowing $N_p$ and $N_q$ ``screens'' the information we obtain from $K_\mathrm{out}$ about $K_\mathrm{in}$, i.e. if we know $N_p$ and $N_q$, learning the value of $K_\mathrm{out}$ gives us no additional information about $K_\mathrm{in}$.

We can now write
\begin{equation}
    P(k_{\mathrm{out}},k_{\mathrm{in}}) = 
    \sum_{N_p,N_q} P(k_\mathrm{out},k_\mathrm{in}|N_p,N_q)
    P(N_p,N_q) \,,    
\end{equation}
where
\begin{equation}
    P(N_p,N_q) = \frac{1}{2^{N-1}}
    \binom{N-1}{N_p}\delta_{N_p+N_q,N-1}
\end{equation}
denotes the probability that a randomly chosen node will be in relation to $N_p$ type-$p$ nodes and $N_q$ type-$q$ nodes. The Kronecker delta, $\delta_{i,j}$, in this expression imposes the constraint that $N_p+N_q = N - 1$. 

To proceed, we will need the probabilities $P(k_\mathrm{out}|N_p,N_q)$ and $P(k_\mathrm{in}|N_p,N_q)$. These probabilities are simply the convolutions
\begin{multline}
    P(k_\mathrm{out}|N_p,N_q) = 
    \sum_{j=0}^{k_\mathrm{out}}
    \left[\binom{N_p}{j}p^j(1-p)^{N_p-j}\right]\times \\
    \times
    \left[\binom{N_q}{k_\mathrm{out}-j}q^{k_\mathrm{out}-j}
    (1-q)^{N_q-(k_\mathrm{out}-j)}\right]
    \label{eq:P k out given Np Nq}
\end{multline}
and
\begin{multline}
    P(k_\mathrm{in}|N_p,N_q) = 
    \sum_{j=0}^{k_\mathrm{in}}
    \left[\binom{N_p}{j}q^j(1-q)^{N_p-j}\right]\times \\
    \times
    \left[\binom{N_q}{k_\mathrm{in}-j}p^{k_\mathrm{in}-j}
    (1-p)^{N_q-(k_\mathrm{in}-j)}\right] \,.
    \label{eq:P k in given Np Nq}
\end{multline}
It will be useful to have the generating functions associated with these distributions. These can be defined as
\begin{align}
    G_\mathrm{out}(x;N_p,N_q) & = \sum_{k_\mathrm{out}}P(k_\mathrm{out}|N_p,N_q)
    x^{k_\mathrm{out}} \,, \\ 
    G_\mathrm{in}(y;N_p,N_q) & = \sum_{k_\mathrm{in}}P(k_{\mathrm{in}}|N_p,N_q)
    y^{k_\mathrm{in}} \,.
\end{align}
Upon substitution, one can carry out the sums without too much effort to obtain
\begin{align}
    G_\mathrm{out}(x;N_p,N_q) & = 
    (1-q+qx)^{N_q}(1-p+px)^{N_p} \,, 
    \label{eq:G out given Np Nq}\\ 
    G_\mathrm{in}(y;N_p,N_q) & = 
    (1-p+py)^{N_q}(1-q+qy)^{N_p} \,.
    \label{eq:G in given Np Nq}
\end{align}

If we now define the generating function of the joint distribution
\begin{equation}
    G(x,y) = \sum_{k_\mathrm{out}}\sum_{k_\mathrm{in}} P(k_{\mathrm{out}},k_{\mathrm{in}})
    x^{k_\mathrm{out}}y^{k_\mathrm{in}} \,,
    \label{eq:G(x,y) def}
\end{equation}
then we find
\begin{multline}
    G(x,y) = \sum_{N_p,N_q} P(N_p,N_q) \times\\ \times 
    \left[\sum_{k_\mathrm{out}} P(k_\mathrm{out}|N_p,N_q)
    x^{k_\mathrm{out}}\right] \times\\ \times 
    \left[\sum_{k_\mathrm{in}} P(k_\mathrm{in}|N_p,N_q)
    y^{k_\mathrm{in}}\right] \,,
\end{multline}
i.e.
\begin{multline}
    G(x,y) = \sum_{N_p,N_q} P(N_p,N_q) \times\\ \times 
    G_\mathrm{out}(x;N_p,N_q)
    G_\mathrm{in}(y;N_p,N_q) \,,
\end{multline}
or explicitly
\begin{multline}
    G(x,y) = \frac{1}{2^{N-1}} \sum_{N_p,N_q} 
    \binom{N-1}{N_p}\delta_{N_p+N_q,N-1} \times\\ \times 
    [(1-p+px)(1-q+qy)]^{N_p} \times\\ \times 
    [(1-q+qx)(1-p+py)]^{N_q} \,.
\end{multline}

This sum is easily calculated using the binomial theorem such that upon simplification, we obtain the beautifully concise expression
\begin{multline}
    G(x,y) = \\
    \left[(1-p+px)(1-q+qy)-\frac{p-q}{2}(x-y)\right]^{N-1} \,.
    \label{eq:intransitive gen func}
\end{multline}
To determine the underlying joint distribution $P(k_{\mathrm{out}},k_{\mathrm{in}})$, we can expand this expression using the multinomial theorem to obtain
\begin{multline}
    G(x,y) = 
    \sum_{k_\mathrm{out},k_\mathrm{in}} \sum_{\ell,j} 
    \binom{N-1}{\ell,j,k_\mathrm{out}-\ell,k_\mathrm{in}-\ell}
    (pq)^\ell
    \times \\ \times
    (1-p)^j(1-q)^j
    \left(\frac{p+q}{2}-pq\right) ^{k_\mathrm{out}+k_\mathrm{in}-2\ell}  x^{k_\mathrm{out}}y^{k_\mathrm{in}}
\end{multline}
where 
\begin{equation}
    \binom{N}{j_0,j_1,j_2,...} = \frac{N!}{j_0!j_1!j_2!...}
\end{equation}
denotes the multinomial coefficient which is assumed to vanish if any of the $j_n$ are negative or if $j_0 + j_1 + j_2 +... \ne N$. Accordingly, we find that
\begin{multline}
    P(k_\mathrm{out},k_\mathrm{in}) = \sum_{\ell,j} 
    \binom{N-1}{\ell,j,k_\mathrm{out}-\ell,k_\mathrm{in}-\ell}
    (pq)^\ell 
    \times \\ \times
    (1-p)^j(1-q)^j
    \left(\frac{p+q}{2}-pq\right)^{k_\mathrm{out}+k_\mathrm{in}-2\ell} \,.
\end{multline}
This expression can easily be summed over $j$ in which case we obtain
\begin{multline}
    P(k_\mathrm{out},k_\mathrm{in}) = \\
    [(1-p)(1-q)]^{N-1}
    \left[\frac{\frac{p+q}{2}-pq}{(1-p)(1-q)}\right]
    ^{k_\mathrm{out}+k_\mathrm{in}}
    \times\\\times
    \sum_{\ell}
    \binom{N-1}{\ell,k_\mathrm{out}-\ell,k_\mathrm{in}-\ell,N-1-k_\mathrm{out}-k_\mathrm{in}+\ell}
    \times\\\times
    \left[\frac{(1-p)(1-q)pq}
    {\left(\frac{p+q}{2}-pq\right)^2}\right]
    ^{\ell} \,,
\end{multline}
or alternatively, in terms of the hypergeometric function $\,_2F_1(a,b;c;z)$ \cite{NIST:DLMF}
\begin{multline}
    P(k_\mathrm{out},k_\mathrm{in}) = \\
    [(1-p)(1-q)]^{N-1}
    \left[\frac{\frac{p+q}{2}-pq}{(1-p)(1-q)}\right]
    ^{k_\mathrm{out}+k_\mathrm{in}}
    \times\\\times
    \binom{N-1}{N-1-k_\mathrm{out}-k_\mathrm{in},k_\mathrm{out},k_\mathrm{in}}
    \times\\\times
    \,_2F_1\left(-k_\mathrm{out},-k_\mathrm{in};
    N-k_\mathrm{out}-k_\mathrm{in};
    \frac{(1-p)(1-q)pq}{\left(\frac{p+q}{2}-pq\right)^2}\right) \,.
\end{multline}

Perhaps more informatively, we can calculate the mixed second moment $\left<K_\mathrm{out}K_\mathrm{in}\right>$ directly from the generating function using the relationship
\begin{equation}
    \left< K_\mathrm{out}K_\mathrm{in}\right>
    = \left.
    \frac{G\left(x,y\right)}{\partial x\partial y} \right|_{x=y=1} \,,
\end{equation}
which gives
\begin{equation}
    \left< K_\mathrm{out}K_\mathrm{in}\right> 
    = (N-1)(N-2)\left(\frac{p+q}{2}\right)^2+(N-1)pq
\end{equation}
such that the covariance of the joint distribution is
\begin{align}
    \mathrm{Cov}\left[K_\mathrm{out},K_\mathrm{in}\right] 
    &= \left< K_\mathrm{out}K_\mathrm{in}\right> - 
    \left< K_\mathrm{out}\right> 
    \left< K_\mathrm{in}\right> \,, \\
    &= -(N-1)\left[\left(\frac{p+q}{2}\right)^2-pq\right] \,,
    \label{eq:intransitive covariance}
\end{align}
where we have made use of Eq.~(\ref{eq:degree statistics intransitive}) for $\left<K_\mathrm{out}\right>$ and $\left<K_\mathrm{in}\right>$. Since $(p+q)/2 \ge \sqrt{pq}$ (the inequality of arithmetic and geometric means), this result states that the out-degree and in-degree of a randomly chosen node are \textit{anti-correlated}. If $p$ and $q$ are taken to zero in the large $N$ limit according to Eqs.~(\ref{eq:lambda_p def}) and (\ref{eq:lambda_q def}), the magnitude of this anti-correlation decays to zero with increasing $N$.

Indeed, in the large $N$ limit, the generating function $G(x,y)$ given by Eq.~(\ref{eq:intransitive gen func}) becomes
\begin{equation}
    G(x,y) \xrightarrow{N\rightarrow\infty} e^{\frac{\lambda_p+\lambda_q}{2}(x-1)} e^{\frac{\lambda_p+\lambda_q}{2}(y-1)} \,.
    \label{eq:G(x,y) intransitive}
\end{equation}
Since the generating function has factored into the product of a function of $x$ and a function of $y$, we find that in the large $N$ limit, the out-degree and in-degree become independent and thus
\begin{multline}
    P(K_\mathrm{out}=k_\mathrm{out},K_\mathrm{in} = k_\mathrm{in}) \xrightarrow{N\rightarrow\infty}  \\
    P(K_\mathrm{out}=k_\mathrm{out})
    P(K_\mathrm{in} = k_\mathrm{in}) 
\end{multline}
where the marginal distributions $P(K_\mathrm{out}=k_\mathrm{out})$ and $P(K_\mathrm{in}=k_\mathrm{in})$ are given by Eq.~(\ref{eq:P k out in large N}). This fits with the intuition we developed above which required $p$ to be large and $q$ to be small. If both $p$ and $q$ are small, a randomly selected node will only connect to a small fraction of other nodes in the network and thus knowing the node's out-degree will only inform us about a small number of other nodes in the network. Since the in-degree will mostly be determined by the behaviour of the large fraction we know nothing about, learning the out-degree of a particular node provides increasingly little information about the in-degree as the size of the network increases.

\subsection{\label{sec:Transitive Degree Distribution}Transitive Networks}

For the transitive nonreciprocal random network, defined by Eq.~(\ref{eq:P_ij transitive}), each node $i$ will have its own out-degree and in-degree distributions $P_i(K_\mathrm{out} = k_\mathrm{out})$ and $P_i(K_\mathrm{in} = k_\mathrm{in})$. The fact that there is no degree distribution for the network as a whole is perhaps the starkest difference between structured and unstructured networks.

In Sec.~\ref{sec:Intransitive Degree Distribution}, we obtained expressions for the out-degree and in-degree distributions of a node in relation to $N_p$ type-$p$ nodes and $N_q$ type-$q$ nodes. The probabilities $P(k_\mathrm{out}|N_p,N_q)$ and $P(k_\mathrm{in}|N_p,N_q)$ are given by Eqs.~(\ref{eq:P k out given Np Nq}) and (\ref{eq:P k in given Np Nq}) while their generating functions, $G_\mathrm{out}(x)$ and 
$G_\mathrm{in}(y)$, are given by Eqs.~(\ref{eq:G out given Np Nq}) and (\ref{eq:G in given Np Nq}). In that context, the quantities $N_p$ and $N_q$ were random variables and thus they needed to be averaged over in order to obtain expressions for the out-degree and in-degree distributions. For the transitive structured network, knowing the identity of node $i$ fully determines $N_p$ and $N_q$ thus we can immediately write
\begin{align}
    P_i(k_\mathrm{out}) &= 
    P(k_\mathrm{out}|N_p=N-i,N_q=i-1) \,, \\
    P_i(k_\mathrm{in}) &= 
    P(k_\mathrm{in}|N_p=N-i,N_q=i-1) \,,
\end{align}
i.e.
\begin{multline}
    P_i(k_\mathrm{out}) = 
    \sum_{j=0}^{k_\mathrm{out}}
    \left[\binom{N-i}{j}p^j(1-p)^{N-i-j}\right]\times \\
    \times
    \left[\binom{i-1}{k_\mathrm{out}-j}q^{k_\mathrm{out}-j}
    (1-q)^{i-1-(k_\mathrm{out}-j)}\right]
\end{multline}
and
\begin{multline}
    P_i(k_\mathrm{in}) = 
    \sum_{j=0}^{k_\mathrm{in}}
    \left[\binom{N-i}{j}q^j(1-q)^{N-i-j}\right]\times \\
    \times
    \left[\binom{i-1}{k_\mathrm{in}-j}p^{k_\mathrm{in}-j}
    (1-p)^{i-1-(k_\mathrm{in}-j)}\right] \,.
\end{multline}
In principle, these sums can be written in terms of the hypergeometric function, $\,_2F_1(a,b;c;z)$, or equivalently, the Jacobi polynomials, $P^{(\alpha,\beta)}_n(x)$ \cite{NIST:DLMF}. In practice, this provides little additional insight beyond the explicit sums given here. 

Instead, it is much more useful to define the generating functions of these distributions
\begin{align}
    G_{\mathrm{out},i}(x) &= \sum_{k_\mathrm{out}} P_i(k_\mathrm{out})x^{k_\mathrm{out}} \,, \\
    G_{\mathrm{in},i}(y) &= \sum_{k_\mathrm{in}} P_i(k_\mathrm{in})y^{k_\mathrm{in}} \,, 
\end{align}
which after subbing in Eqs.~(\ref{eq:G out given Np Nq}) and (\ref{eq:G in given Np Nq}) read
\begin{align}
    G_{\mathrm{out},i}(x) &= 
    (1-q+qx)^{i-1}(1-p+px)^{N-i} \,, \\
    G_{\mathrm{in},i}(y) &= 
    (1-p+py)^{i-1}(1-q+qy)^{N-i} \,. 
\end{align}

Let
\begin{equation}
    \alpha = \frac{i-1}{N-1}\in [0,1] \label{eq:alpha def}
\end{equation}
denote an alternative index to $i$ which denotes the fraction of nodes ``less than'' node $i$ in the network. This is useful because in the large $N$ limit, this index becomes continuous. Then holding $\lambda_p = (N-1)p$ and $\lambda_q = (N-1)q$ constant, we obtain
\begin{align}
    G_{\mathrm{out},\alpha}(x) 
    &\xrightarrow{N\rightarrow\infty} 
    e^{[(1-\alpha)\lambda_p+\alpha\lambda_q](x-1)} \,,\\
    G_{\mathrm{in},\alpha}(y) 
    &\xrightarrow{N\rightarrow\infty}
    e^{[\alpha\lambda_p+(1-\alpha)\lambda_q](y-1)} \,.
\end{align}
Since these are simply the generating functions of Poisson distributions, we find that in the large $N$ limit
\begin{align}
    P_\alpha(k_\mathrm{out}) &= 
    \frac{[(1-\alpha)\lambda_p+\alpha\lambda_q] 
    ^{k_\mathrm{out}}}{k_\mathrm{out}!} 
    e^{-[(1-\alpha)\lambda_p+\alpha\lambda_q]} \,, 
    \label{eq:P alpha k out}\\ 
    P_\alpha(k_\mathrm{in}) &= 
    \frac{[\alpha\lambda_p+(1-\alpha)\lambda_q]
    ^{k_\mathrm{in}}}{k_\mathrm{in}!}
    e^{-[\alpha\lambda_p+(1-\alpha)\lambda_q]} \,,
    \label{eq:P alpha k in}
\end{align}
i.e. the out-degree and in-degree distributions of each node are just Poisson distributions $\mathrm{Pois}(\lambda)$ with a parameter $\lambda$ that varies linearly between $\lambda_p$ and $\lambda_q$. Accordingly
\begin{align}
    \left<K_{\mathrm{out},\alpha}\right> & =
    \mathrm{Var}[K_{\mathrm{out},\alpha}] = 
    (1-\alpha)\lambda_p+\alpha\lambda_q \,, 
    \label{eq:transitive mean and var vs alpha}\\
    \left<K_{\mathrm{in},\alpha}\right> & =
    \mathrm{Var}[K_{\mathrm{in},\alpha}] = 
    \alpha\lambda_p+(1-\alpha)\lambda_q \,.
\end{align}

In Sec.~\ref{sec:Intransitive Degree Distribution}, we saw that the joint degree distribution $P(k_\mathrm{out},k_\mathrm{in})$ of a completely unstructured nonreciprocal random network does not simply factor because the out-degree and in-degree are not necessarily independent (though this independence is obtained in the large $N$ limit). We also saw however that if we condition on the number of type-$p$ and type-$q$ nodes, $N_p$ and $N_q$, that independence does hold. In contrast, for the transitive nonreciprocal random network, the degree distributions are necessarily conditioned over $N_p$ and $N_q$. Accordingly, for the transitive case, the out-degree and in-degree of any particular node are always independent and thus we find that for networks of \textit{any} size
\begin{equation}
    P_i(k_\mathrm{out},k_\mathrm{in}) = 
    P_i(k_\mathrm{out})P_i(k_\mathrm{in}) \,.
\end{equation}

Though the Poisson distributions obtained in Eqs.~(\ref{eq:P alpha k out}) and (\ref{eq:P alpha k in}) superficially resemble the Poisson distribution obtained for the unstructured nonreciprocal random network, given in Eq.~(\ref{eq:P k out in large N}), there are crucial differences reflecting the difference between the structured network and unstructured one. For instance, the dependence on $\alpha$ in Eqs.~(\ref{eq:P alpha k out}) and (\ref{eq:P alpha k in}) means that for almost all nodes, $P_\alpha(k_\mathrm{out})\ne P_\alpha(k_\mathrm{in})$. Further, if we try to define an out-degree or in-degree distribution for a typical node by say, averaging over $\alpha$, we obtain
\begin{align}
    \overline{P_\alpha(k_\mathrm{out})} 
    &= \int_0^1d\alpha\,P_\alpha(k_\mathrm{out}) \,, \\
    &= \frac{1}{\lambda_p-\lambda_q}
    \int_{\lambda_q}^{\lambda_p}dt\,
    \frac{1}{k_\mathrm{out}!}t^{k_\mathrm{out}}e^{-t} \,,\\
    &= \frac{1}{k_\mathrm{out}!} 
    \frac{\Gamma(k_\mathrm{out}+1,\lambda_q)-
    \Gamma(k_\mathrm{out}+1,\lambda_p)}
    {\lambda_p-\lambda_q} \,,
\end{align}
and the same for $\overline{P_\alpha(k_\mathrm{in})}$. Here $\Gamma(a,z)$ denotes the incomplete Gamma function \cite{NIST:DLMF}. This distribution is a mixed Poisson-uniform distribution \cite{Grandell1997Book} and in general, it is substantially different from the Poisson distribution. For instance, we can calculate its factorial moments fairly readily
as
\begin{align}
    \left< \, \overline{ \frac{K_\mathrm{out}!}
    {(K_\mathrm{out}-n)!} }\, \right>
    &= \sum_{k_\mathrm{out}}
    \frac{k_\mathrm{out}!}{(k_\mathrm{out}-n)!}
    \overline{P_{\alpha}(k_\mathrm{out})} \,, \\
    &= \frac{1}{n+1}
    \frac{\lambda_p^{n+1}-\lambda_q^{n+1}}
    {\lambda_p-\lambda_q} \,.
\end{align}
Here, we have introduced the notation that $\overline{f(K_\mathrm{out})}$ denotes an average calculated over the nodes in a particular network realisation and $\left< f(K_\mathrm{out}) \right>$ denotes an average over the ensemble of networks. If $\lambda_p$ and $\lambda_q$ are close to each other such that we can write $\lambda_p = (\lambda_p+\lambda_q)/2 + \delta$ and $\lambda_q = (\lambda_p+\lambda_q)/2 - \delta$ where $\delta=(\lambda_p-\lambda_q)/2\ll 1$, then this expression becomes
\begin{equation}
    \left< \, \overline{ \frac{K_\mathrm{out}!}
    {(K_\mathrm{out}-n)!} }\, \right> = 
    \left(\frac{\lambda_p+\lambda_q}{2}\right)^n
    [1+O(\delta)] \,.
\end{equation}
Since the $n^\text{th}$-factorial moment of an ordinary Poisson distribution $\mathrm{Pois}(\lambda)$ is $\lambda^n$, we find that the mixed Poisson-uniform distribution will only resemble the Poisson distribution given by Eq.~(\ref{eq:P k out in large N}) if $\delta$ is small, i.e. if $\lambda_p$ and $\lambda_q$ are close to each other. In particular, the mean and variance of the mixed Poisson-uniform distribution are
\begin{align}
    \left\langle \overline{K_\mathrm{out}}\right\rangle 
    &=\frac{\lambda_p+\lambda_q}{2} \,,
    \label{eq:transitive mean degree}\\
    \overline{\mathrm{Var}}\left[K_\mathrm{out}\right] 
    &= \frac{\lambda_p+\lambda_q}{2}+
    \frac{(\lambda_p-\lambda_q)^{2}}{12} \,,
    \label{eq:transitive variance}
\end{align}
where we have denoted $\overline{\mathrm{Var}}\left[K_\mathrm{out}\right] = \left< \overline{K_\mathrm{out}^2} - \overline{K_\mathrm{out}}^2 \right>$ as the ensemble average of the network out-degree variance.
Accordingly, even a so-called ``typical node'' in the structured network doesn't generically resemble a typical node in the unstructured network.

In the same fashion, we can consider the joint distribution of a ``typical'' node in the structured network
\begin{equation}
    \overline{P_\alpha(k_\mathrm{out},k_\mathrm{in})} = 
    \int_0^1d\alpha 
    P_\alpha(k_\mathrm{out})P_\alpha(k_\mathrm{in}) \,,
\end{equation}
which can be written in terms of the incomplete Beta function $\mathrm{B}_z(n,m)$ \cite{NIST:DLMF} as
\begin{multline}
    \overline{P_\alpha(k_\mathrm{out},k_\mathrm{in})} = \frac{(\lambda_p+\lambda_q)^
    {k_\mathrm{out}+k_\mathrm{in}+1}
    e^{-(\lambda_p+\lambda_q)}}
    {(\lambda_q-\lambda_p)k_\mathrm{out}!k_\mathrm{in}!}
    \times\\\times
    [\mathrm{B}_{\lambda_q/(\lambda_p+\lambda_q)}
    (k_\mathrm{out}+1,k_\mathrm{in}+1) \\
    - \mathrm{B}_{\lambda_p/(\lambda_p+\lambda_q)}
    (k_\mathrm{out}+1,k_\mathrm{in}+1)] \,.
    \label{eq:joint deg dist typical}
\end{multline}
Note that this doesn't factor, i.e. even though the out-degree and in-degree of each individual node are independent, the out-degree and in-degree of a hypothetical representative node are not. Specifically, the covariance can easily be extracted from the joint generating function
\begin{align}
    \overline{G_\alpha(x,y)} &= 
    \int_0^1d\alpha\,G_\alpha(x,y) \, \\
    &= \frac{e^{\lambda_p(y-1)+\lambda_q(x-1)}-
    e^{\lambda_p(x-1)+\lambda_q(y-1)}}
    {(\lambda_p-\lambda_q)(y-x)} \,, 
    \label{eq:G(x,y) transitive}
\end{align}
as
\begin{align}
    \overline{\mathrm{Cov}}[K_\mathrm{out},K_\mathrm{in}] &= 
    \left\langle 
    \overline{K_\mathrm{out}K_\mathrm{in}} 
    -\overline{K_\mathrm{out}}\,\overline{K_\mathrm{in}}
    \right\rangle \,,\\
    & = -\frac{(\lambda_p-\lambda_q)^2}{12} \,.
    \label{eq:transitive covariance}
\end{align}
Note that this is an ensemble average of the degree covariance of a so called ``typical node'' obtained by averaging over a network realisation. In contrast, since the out-degree and in-degree of each individual node are independent, the network average of the ensemble covariance vanishes
\begin{multline}
    \overline{\mathrm{Cov}[K_\mathrm{out,\alpha },K_\mathrm{in,\alpha }]} =\\ 
    \overline{\left< K_\mathrm{out,\alpha }K_\mathrm{in,\alpha } \right> -
    \left< K_\mathrm{out,\alpha } \right>\left< K_\mathrm{in,\alpha } \right>} 
    = 0 \,.
    \label{eq:transitive covariance 2}
\end{multline}

In general, one of the characteristic features of structured networks is that functions of higher order moments such as the variance and covariance do not satisfy commutativity of ensemble and network averaging, i.e. \mbox{$\overline{\mathrm{Var}[K_\mathrm{out,\alpha}]} \ne \overline{\mathrm{Var}}[K_\mathrm{out}]$}, \mbox{$\overline{\mathrm{Cov}[K_\mathrm{out,\alpha},K_\mathrm{in,\alpha}]} \ne \overline{\mathrm{Cov}}[K_\mathrm{out},K_\mathrm{in}]$}, etc.

\section{\label{sec:Percolation}Percolation}

As discussed in the introduction, the standard techniques for studying percolation are of limited use when dealing with structured networks. Accordingly, we will develop an alternative approach based on determining self-consistent equations for the probability that a node $i$ can reach either some other node $j$ or the giant component of the network.

To investigate the percolation properties of nonreciprocal random networks, we treat the unpercolated and percolated states separately. Let $P_{i\rightarrowtriangle j}$ denote the probability that there exists a path connecting node $i$ to some other node $j$. If the network is locally tree-like beneath percolation, since there are few paths connecting any given pair of nodes, we can self-consistently write at lowest order
\begin{equation}
    P_{i\rightarrowtriangle j} \approx P_{i\rightarrow j} + 
    \sum_{k\ne i,j}P_{i\rightarrow k}P_{k\rightarrowtriangle j} 
    \,. \label{eq:subpercolation}
\end{equation}
This equation asserts that beneath percolation, the probability that node $i$ can reach node $j$ is just equal to the probability that a direct connection exists between $i$ and $j$ plus the probability that a path whose first step is via some intermediate node $k$ exists ($\forall k\ne i,j$). The major advantage of this formalism is that Eq.~(\ref{eq:subpercolation}) is linear in $P_{i\rightarrowtriangle j}$ and thus amenable to solution by standard techniques. Higher order nonlinear corrections would have to account for the possibility that multiple such paths exist and thus the right-hand side of Eq.~(\ref{eq:subpercolation}) is technically an overestimate of $P_{i\rightarrowtriangle j}$.

For the large $N$ limit, in a similar manner to Eq.~(\ref{eq:alpha def}), it is helpful to define the indices
\begin{align}
    \alpha &= \frac{i-1}{N-1} \in [0,1] \,,\\
    \beta &= \frac{j-1}{N-1} \in [0,1] \,,\\
    \gamma &= \frac{k-1}{N-1} \in [0,1] \,,
\end{align}
which can be considered continuous. We also define the probability densities
\begin{align}
    f_{\alpha\rightarrow\beta} &= 
    (N-1)P_{i\rightarrow j} \,, \\
    f_{\alpha\rightarrowtriangle\beta} &= 
    (N-1)P_{i\rightarrowtriangle j} \,,
\end{align}
in which case Eq.~(\ref{eq:subpercolation}) becomes the linear integral equation
\begin{equation}
    f_{\alpha\rightarrowtriangle\beta} \approx 
    f_{\alpha\rightarrow\beta} + 
    \int_0^1d\gamma\, f_{\alpha\rightarrow\gamma} f_{\gamma\rightarrowtriangle\beta} \,.
    \label{eq:gen integ eq}
\end{equation}

One can use the solution to this integral equation to calculate a variety of other quantities. For instance, the mean number of nodes which can be reached from node $i$, $\left< N_{O,i}\right>$, or the mean number of nodes which can reach node $j$, $\left< N_{I,j}\right>$, are simply
\begin{align}
    \left< N_{O,i}\right> &= 
    \sum_{j\ne i}P_{i\rightarrowtriangle j} 
    = \int_0^1 d\beta f_{\alpha\rightarrowtriangle \beta} \,, 
    \label{eq:N_O}\\
    \left< N_{I,j}\right> &= 
    \sum_{i\ne j}P_{i\rightarrowtriangle j} 
    = \int_0^1 d\alpha f_{\alpha\rightarrowtriangle \beta} \,.
    \label{eq:N_I}
\end{align}
When these quantities are order $O(N)$, the percolation transition has been reached and Eq.~(\ref{eq:subpercolation}) ceases to be valid.

Above the percolation threshold, a giant strongly connected component (GSCC) emerges in the network. This component contains a macroscopic fraction $S$ of the network's nodes and is characterised by the property that for any two nodes $i$ and $j$ in the GSCC, there exists a path from $i$ to $j$. In addition to the GSCC, one can define two giant components, the giant out-component (GOC), consisting of fraction $O$ of the network, and the giant in-component (GIC), consisting of fraction $I$ of the network. A node $i$ belongs to the GOC if it can be reached from a node in the GSCC. Similarly, a node $i$ belongs to the GIC if it can reach nodes in the GSCC. Note that according to these definitions, nodes in the GSCC belong to both the GOC and the GIC and in fact, the GSCC is precisely the intersection of the GOC and GIC \cite{Dorogovtsev2001}.

For any node $\alpha = (i-1)/(N-1)$, let $P_S(\alpha)$, $P_O(\alpha)$ and $P_I(\alpha)$ denote the probabilities that node $\alpha$ is in the GSCC, GOC and GIC respectively. For each of these functions, we can define a corresponding cumulative probability function
\begin{align}
    F_S(\alpha) &= \int_0^\alpha d\gamma \, P_S(\gamma) 
    & F_S'(\alpha) &= P_S(\alpha) \,, \\
    F_O(\alpha) &= \int_0^\alpha d\gamma \, P_O(\gamma) 
    & F_O'(\alpha) &= P_O(\alpha) \,, \\
    F_I(\alpha) &= \int_0^\alpha d\gamma \, P_I(\gamma) 
    & F_I'(\alpha) &= P_I(\alpha) \,.
\end{align}
The cumulative probabilities are characterised by the boundary values
\begin{align}
    F_S(0) &= 0 & F_S(1) &= S \label{eq:BC 1}\\
    F_O(0) &= 0 & F_O(1) &= O \label{eq:BC 2}\\
    F_I(0) &= 0 & F_I(1) &= I \label{eq:BC 3}
\end{align}
To determine these functions, we imagine an instantiated network of $N-1$ nodes to which we add a single new node $\alpha$. In the large $N$ limit, adding a single node has no effect on these macroscopic quantities and thus we can self-consistently write that the new node $\alpha$ will belong to the GSCC with probability
\begin{equation}
    P_S(\alpha) = 
    P(O\rightarrow\alpha, \alpha \rightarrow I) \,,
    \label{eq:ODE 1}
\end{equation}
where $P(O\rightarrow\alpha, \alpha \rightarrow I)$ denotes the probability that the new node will be connected to from at least one node in the GOC and will connect to at least one node in the GIC. Similarly, the new node $\alpha$ will belong to the GOC or GIC with probabilities 
\begin{align}
    P_O(\alpha) &= P(O\rightarrow\alpha) \,, 
    \label{eq:ODE 2}\\
    P_I(\alpha) &= 
    P(\alpha \rightarrow I) \,.
    \label{eq:ODE 3}
\end{align}
If the probabilities $P(\alpha \rightarrow I)$ and $P(O\rightarrow\alpha)$ can be written in terms of the cumulative probabilities $F_S(\alpha)$, $F_O(\alpha)$ and $F_I(\alpha)$, then these three equations will define coupled, first-order, ordinary differential equations (ODEs) for these functions. The solution to these equations will then determine the sizes of the GSCC, GOC and GIC. We will see how this works concretely in the following subsections.

\subsection{Intransitive Networks}

As discussed in the introduction, the percolation properties of an unstructured network can be treated with standard methods such as the generating function method \cite{Newman2001, Dorogovtsev2001, Schwartz2002, Boguna2005, Meyers2006}. Since these methods fail for structured networks, we will apply the technique just described to the unstructured case as well. This will help develop familiarity with the method while also demonstrating its full generality. The agreement for unstructured networks between this method and the generating function method is demonstrated in Appendix~\ref{sec:GF Method}.

Begin by supposing we are in the unpercolated state. For the intransitive nonreciprocal random network, \mbox{$P_{\alpha\rightarrow\beta} = (p+q)/2 = \frac{\lambda_p+\lambda_q}{N-1}/2$} thus Eq.~(\ref{eq:gen integ eq}) reads
\begin{equation}
    f_{\alpha\rightarrowtriangle\beta} \approx 
    \frac{\lambda_p+\lambda_q}{2} + 
    \frac{\lambda_p+\lambda_q}{2}
    \int_0^1d\gamma\,f_{\gamma\rightarrowtriangle\beta} \,.
\end{equation}
This is an integral equation for $f_{\alpha\rightarrowtriangle\beta}$. Since all nodes are statistically equivalent in the intransitive network, the probability that any node $\alpha$ can reach any other node $\beta$ cannot depend on either $\alpha$ or $\beta$ and thus we can immediately guess that $f_{\alpha\rightarrowtriangle\beta} = c$ is a solution for some constant $c$. Substitution into the integral equation thus gives
\begin{equation}
    f_{\alpha\rightarrowtriangle\beta} \approx 
    \frac{\lambda_p+\lambda_q}{2-(\lambda_p+\lambda_q)}
     \,.
\end{equation}
Since $f_{\alpha\rightarrowtriangle\beta} \ge 0$, this expression is only valid when
\begin{equation}
    \lambda_p+\lambda_q < 2 \,,
\end{equation}
i.e. percolation occurs when 
\begin{equation}
    \lambda_p+\lambda_q = 2.
    \label{eq:percolation line intransitive}
\end{equation}

Using Eqs.~(\ref{eq:N_O}) and (\ref{eq:N_I}), we can determine that the mean number of nodes which can be reached from a typical node or reach a typical node are just
\begin{equation}
    \left< N_O \right> = \left< N_I \right> = 
    \frac{\lambda_p+\lambda_q}{2-(\lambda_p+\lambda_q)} \,.
\end{equation}
Accordingly, for large but finite networks of size $N$, the assumptions we have been using will break down when
\begin{equation}
    \frac{\lambda_p+\lambda_q}{2-(\lambda_p+\lambda_q)} 
    \sim O(N) \,.
\end{equation}

Let us now move on to the percolated situation. For the intransitive network, we can write
\begin{align}
    P(O\rightarrow \alpha) &= 
    1 - \left(1 - \frac{p+q}{2}\right)^{(N-1)O} \,, \\
    P(\alpha\rightarrow I) &= 
    1 - \left(1 - \frac{p+q}{2}\right)^{(N-1)I} \,, 
\end{align}
thus in the large $N$ limit
\begin{align}
    P(O\rightarrow \alpha) &= 
    1 - e^{-\frac{\lambda_p+\lambda_q}{2}O} \,, \\
    P(\alpha\rightarrow I) &= 
    1 - e^{-\frac{\lambda_p+\lambda_q}{2}I} \,.
\end{align}
Accordingly, Eqs.~(\ref{eq:ODE 1}), (\ref{eq:ODE 2}) and (\ref{eq:ODE 3}) can be written as
\begin{align}
    F_S' &= 
    \left[1 - e^{-\frac{\lambda_p+\lambda_q}{2}O}\right]
    \left[1 - e^{-\frac{\lambda_p+\lambda_q}{2}I}\right] \,, \\
    F_O' &= 1 - e^{-\frac{\lambda_p+\lambda_q}{2}O}  \,, \\
    F_I' &= 1 - e^{-\frac{\lambda_p+\lambda_q}{2}I} \,.
\end{align}
These three equations constitute three coupled ODEs for the functions $F_S(\alpha)$, $F_O(\alpha)$ and $F_I(\alpha)$. Their boundary conditions are given by Eqs.~(\ref{eq:BC 1}), (\ref{eq:BC 2}) and (\ref{eq:BC 3}). Since the right-hand side of these equations don't depend on $\alpha$, they are easily integrated and we obtain
\begin{align}
    F_S &= 
    \left[1 - e^{-\frac{\lambda_p+\lambda_q}{2}O}\right]
    \left[1 - e^{-\frac{\lambda_p+\lambda_q}{2}I}\right]
    \alpha \,, \\
    F_O &= 
    \left[1 - e^{-\frac{\lambda_p+\lambda_q}{2}O}\right]
    \alpha \,, \\
    F_I &= 
    \left[1 - e^{-\frac{\lambda_p+\lambda_q}{2}I}\right]
    \alpha \,,
\end{align}
where we have used the boundary conditions $F_S(0) = F_O(0) = F_I(0) = 0$ to determine the constants of integration for each equation. We can now finally determine $S$, $I$ and $O$ by subbing in $\alpha = 1$. This gives the coupled equations
\begin{align}
    S &= 
    \left[1 - e^{-\frac{\lambda_p+\lambda_q}{2}O}\right]
    \left[1 - e^{-\frac{\lambda_p+\lambda_q}{2}I}\right] \,, 
    \label{eq:S equ intransitive} \\
    O &= 1 - e^{-\frac{\lambda_p+\lambda_q}{2}O} \,, 
    \label{eq:O equ intransitive} \\
    I &= 1 - e^{-\frac{\lambda_p+\lambda_q}{2}I} \,.
    \label{eq:I equ intransitive}
\end{align}
Note that $S = O = I = 0$ is a valid solution to these equations. This corresponds to the unpercolated state. For $\lambda_p +\lambda_q > 2$, we expect an additional solution corresponding to the percolated state.

Conveniently, Eqs.~(\ref{eq:O equ intransitive}) and (\ref{eq:I equ intransitive}) are decoupled and thus can be solved directly to give
\begin{equation}
    O = I = 1 + 
    \frac{2}{\lambda_p +\lambda_q}
    W\left(-\frac{\lambda_p+\lambda_q}{2}
    e^{-\frac{\lambda_p+\lambda_q}{2} } \right) \,,
    \label{eq:O, I intransitive}
\end{equation}
where $W(z)$ denotes the principle branch of the Lambert-$W$ function \cite{NIST:DLMF}. Similarly, comparing Eqs.~(\ref{eq:S equ intransitive}), (\ref{eq:O equ intransitive}) and (\ref{eq:I equ intransitive}), we clearly have $S=OI$ and thus
\begin{equation}
    S = 
    \left[1 + \frac{2}{\lambda_p+\lambda_q}
    W\left(-\frac{\lambda_p+\lambda_q}{2} 
    e^{-\frac{\lambda_p+\lambda_q}{2}}\right)\right]^2 \,.
    \label{eq:S intransitive}
\end{equation}

We have thus determined the sizes of the GSCC, GOC and GIC. Unsurprisingly, the solution is identical to that of an ordinary directed random network with parameter $\lambda = (\lambda_p + \lambda_q)/2$ (see Appendix~\ref{sec:GF Method}). This was to be expected as the intransitive nonreciprocal random network is characterised in the large $N$ limit by identical independent Poissonian out-degree and in-degree distributions and is thus, in this sense, indistinguishable from an ordinary random directed network. Indeed, properties of the solution such as $O = I$ and $S = OI$ could have been predicted in advance on the basis that the out-degree and in-degree distributions are identical and independent (as observed in \cite{Dorogovtsev2001}). We have never-the-less shown explicitly how to derive the solution in this relatively simple case as it will provide guidance for the more complicated transitive case treated below.

\subsection{Transitive Networks}

The percolation properties of the transitive nonreciprocal random network are much more involved than the intransitive one. To see this begin with the situation beneath the percolation threshold. In this case, Eq.~(\ref{eq:gen integ eq}) reads differently depending on whether $\alpha<\beta$ or whether $\alpha>\beta$
\begin{multline}
    f_{\alpha\rightarrowtriangle\beta} = \\
    \begin{cases}
        \lambda_p + \lambda_q\int_0^\alpha d\gamma\, 
        f_{\gamma\rightarrowtriangle\beta} + 
        \lambda_p\int_\alpha^1 d\gamma\, 
        f_{\gamma\rightarrowtriangle\beta} & \alpha<\beta \\
        \lambda_q + \lambda_q\int_0^\alpha d\gamma\, 
        f_{\gamma\rightarrowtriangle\beta} + 
        \lambda_p\int_\alpha^1 d\gamma\, 
        f_{\gamma\rightarrowtriangle\beta} & \alpha>\beta
    \end{cases} \,.
\end{multline}
To solve this integral equation, it will be helpful to differentiate it with respect to $\alpha$ in which case we obtain the ODEs
\begin{equation}
    \frac{\partial f_{\alpha\rightarrowtriangle\beta}}
    {\partial\alpha} = 
    \begin{cases}
        (\lambda_q-\lambda_p)
        f_{\alpha\rightarrowtriangle\beta} & \alpha<\beta \\
        (\lambda_q-\lambda_p)
        f_{\alpha\rightarrowtriangle\beta} & \alpha>\beta
    \end{cases} \,.
\end{equation}
Though these ODEs are identical for both $\alpha<\beta$ and $\alpha>\beta$, the solution in each region can differ by the constant of integration and thus we continue to treat them as if they were two distinct ODEs. Their solutions are just
\begin{equation}
    f_{\alpha\rightarrowtriangle\beta} = 
    \begin{cases}
        C_1(\beta)e^{(\lambda_q-\lambda_p)\alpha} & 
        \alpha<\beta\\
        C_2(\beta)e^{(\lambda_q-\lambda_p)\alpha} & 
        \alpha>\beta
    \end{cases} \,,
\end{equation}  
where $C_1(\beta)$ and $C_2(\beta)$ denote constants of integration which can depend on $\beta$. To determine these functions, we substitute back into the integral equation which results in the pair of equations
\begin{multline}
    0 = \lambda_p(\lambda_q-\lambda_p) + C_1(\beta)
    \left[\lambda_pe^{(\lambda_q-\lambda_p)\beta} 
    - \lambda_q\right] + \\
    + C_2(\beta)\lambda_p\left[e^{\lambda_q-\lambda_p}
    -e^{(\lambda_q-\lambda_p)\beta}\right] 
\end{multline}
and
\begin{multline}
    0 = \lambda_q(\lambda_q-\lambda_p) + C_1(\beta)
    \lambda_q\left[e^{(\lambda_q-\lambda_p)\beta}-1\right] + \\
    + C_2(\beta)\left[\lambda_pe^{\lambda_q-\lambda_p}
    - \lambda_qe^{(\lambda_q-\lambda_p)\beta}\right] \,.
\end{multline}
These equations are easily solved for $C_1(\beta)$ and $C_2(\beta)$
\begin{align}
    C_1(\beta) &= 
    \frac{\lambda_q-\lambda_p}
    {\lambda_q-\lambda_pe^{\lambda_q-\lambda_p}}\lambda_p
    e^{(\lambda_q-\lambda_p)\left(1-\beta\right)} \,, \\
    C_2(\beta) &= 
    \frac{\lambda_q-\lambda_p}
    {\lambda_q-\lambda_pe^{\lambda_q-\lambda_p}}
    \lambda_qe^{-(\lambda_q-\lambda_p)\beta} \,,
\end{align}
such that
\begin{equation}
    f_{\alpha\rightarrowtriangle\beta} = 
    \frac{\lambda_q-\lambda_p}
    {\lambda_q-\lambda_pe^{\lambda_q-\lambda_p}}
    \begin{cases}
        \lambda_pe^{(\lambda_q-\lambda_p)[1-(\beta-\alpha)]}
        & \alpha<\beta\\
        \lambda_qe^{(\lambda_q-\lambda_p)(\alpha-\beta)} 
        & \alpha>\beta
    \end{cases} \,.
    \label{eq:f alpha -> beta transitive}
\end{equation}

This expression is valid so long as $f_{\alpha\rightarrowtriangle\beta} > 0$ and when this ceases to be the case, percolation will occur. To determine the point where percolation sets in, it will be helpful to define the function
\begin{equation}
    C(\lambda_p,\lambda_q) = 
    \frac{\lambda_q-\lambda_pe^{\lambda_q-\lambda_p}}
    {\lambda_q-\lambda_p} \,.
\end{equation}
For convenience, suppose that we hold $\lambda_p$ fixed and vary $\lambda_q$. Though $C(\lambda_p,\lambda_q)$ appears to be ill-defined when $\lambda_q\rightarrow\lambda_p$, application of l'Hopital's rule to this case gives
\begin{equation}
    C(\lambda_p,\lambda_q) 
    \xrightarrow{\lambda_q\rightarrow\lambda_p} 
    1-\lambda_p
\end{equation}
thus it is in fact well-defined for all $\lambda_q$.
Further, we have 
\begin{equation}
    C(\lambda_p,\lambda_q) 
    \xrightarrow{\lambda_q\rightarrow0} 
    e^{-\lambda_p} > 0
\end{equation}
thus for sufficiently small $\lambda_q$, $C(\lambda_p,\lambda_q)>0$. Differentiating $C(\lambda_p,\lambda_q)$ with respect to $\lambda_q$ gives
\begin{multline}
    \frac{\partial}{\partial\lambda_q}C(\lambda_p,\lambda_q)=\\
    \frac{\lambda_pe^{\lambda_q-\lambda_p}}
    {(\lambda_q-\lambda_p)^2}
    \left[[1-(\lambda_q-\lambda_p)]
    -e^{-(\lambda_q-\lambda_p)}\right] \,,
\end{multline}
but for all $x\in\mathbb{R}$, $ 1-x \le e^{-x}$ thus we must have $\partial_{\lambda_q}C(\lambda_p,\lambda_q) \le 0 $ for all $\lambda_q$, i.e. $C(\lambda_p,\lambda_q)$ is a decreasing function of $\lambda_q$. We conclude that $C(\lambda_p,\lambda_q)$ will only be positive for values  $\lambda_q \in [0,\lambda_{q,c})$ where 
$\lambda_{q,c}(\lambda_p)$ is the first root of $C(\lambda_p,\lambda_q)$ (assuming such a root exists), i.e. $\lambda_{q,c}(\lambda_p)$ is defined as the solution to $C(\lambda_p,\lambda_{q,c}(\lambda_p)) = 0$. Explicitly, we have
\begin{equation}
    0 = \lambda_{q,c}-\lambda_pe^{\lambda_{q,c}-\lambda_p} \,,
\end{equation}
which has solution
\begin{equation}
    \lambda_{q,c}(\lambda_p) = 
    \begin{cases}
        -W_{-1}\left(-\lambda_pe^{-\lambda_p}\right) &
        \lambda_p\le1 \\
        -W_0\left(-\lambda_pe^{-\lambda_p}\right) & 
        \lambda_p\ge1 
    \end{cases}\,,
    \label{eq:percolation line transitive}
\end{equation}
where $W_0(x)$ and $W_{-1}(x)$ denote the primary and secondary branches of the Lambert-$W$ function respectively \cite{NIST:DLMF}.

More informatively, we can divide the phase-space $(\lambda_p,\lambda_q)$ into two regions. When
\begin{equation}
    \frac{\lambda_q-\lambda_p}{\ln(\lambda_q/\lambda_p)} < 1 
    \,, \label{eq:transitive unpercolated condition}
\end{equation}
the network will be unpercolated, whereas when
\begin{equation}
    \frac{\lambda_q-\lambda_p}{\ln(\lambda_q/\lambda_p)} > 1 
    \,, \label{eq:transitive percolated condition}
\end{equation}
the network will be percolated. The critical line where percolation occurs is when equality holds.

Assuming we are beneath percolation, we can use our result for $f_{\alpha\rightarrowtriangle\beta}$ given by Eq.~(\ref{eq:f alpha -> beta transitive}) in Eqs.~(\ref{eq:N_O}) and (\ref{eq:N_I}) to determine the average number of nodes which node $\alpha$ can reach and be reached by
\begin{align}
    \left\langle N_{O,\alpha}\right\rangle &= 
    \frac{1}{C(\lambda_p,\lambda_q)}
    e^{(\lambda_q-\lambda_p)\alpha}-1 \,, \\
    \left\langle N_{I,\alpha}\right\rangle &= 
    \frac{1}{C(\lambda_p,\lambda_q)}
    e^{(\lambda_q-\lambda_p)\left(1-\alpha\right)}-1 \,,
\end{align}
For large but finite networks consisting of $N$ nodes, we can use these expressions to determine that percolation will be approached when
\begin{equation}
    \frac{e^{|\lambda_q-\lambda_p|}}{C(\lambda_p,\lambda_q)}
    \sim O(N) \,.
\end{equation}

\begin{figure}
    \centering
    \includegraphics[width=0.49\linewidth]{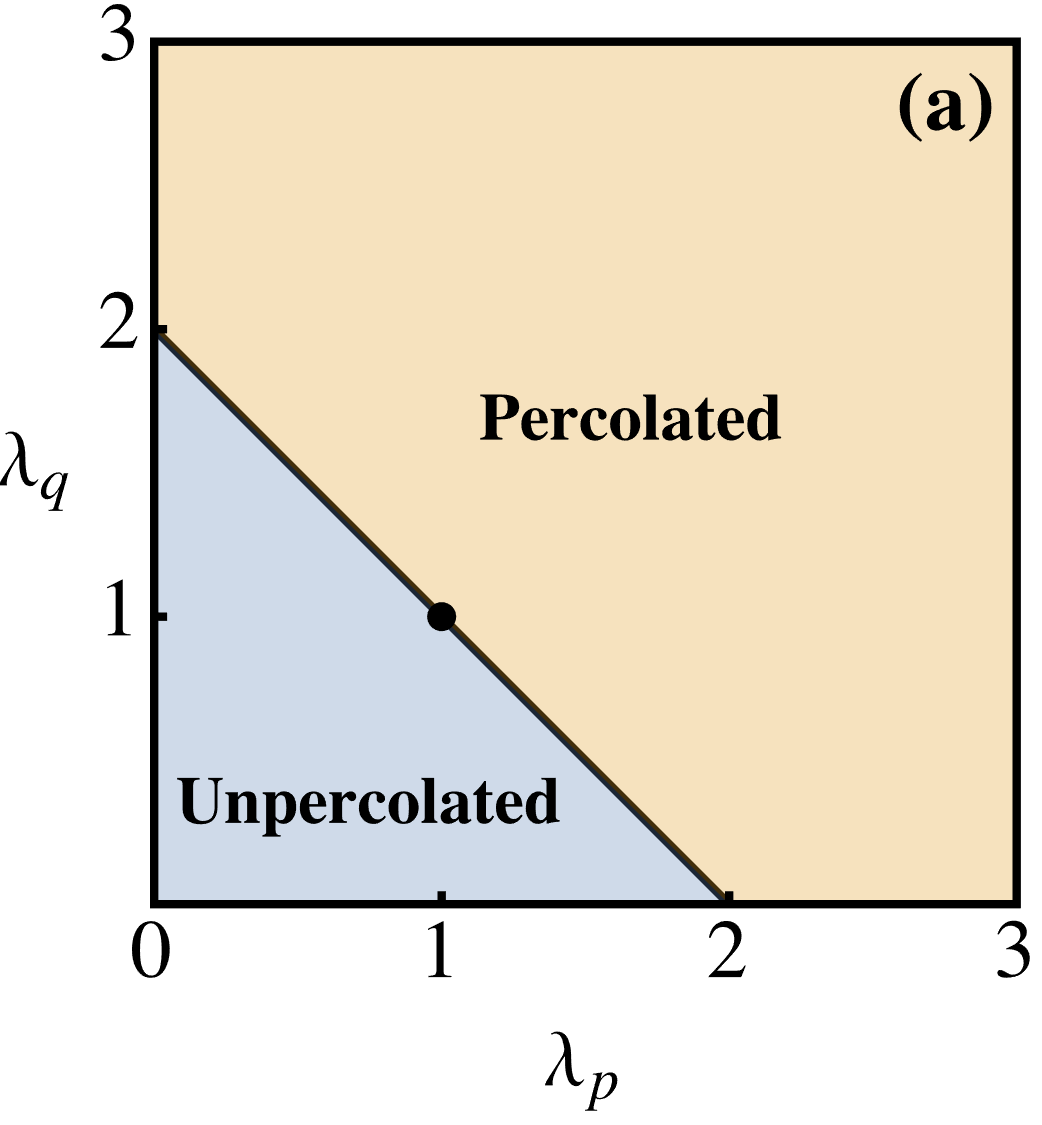}
    \includegraphics[width=0.49\linewidth]{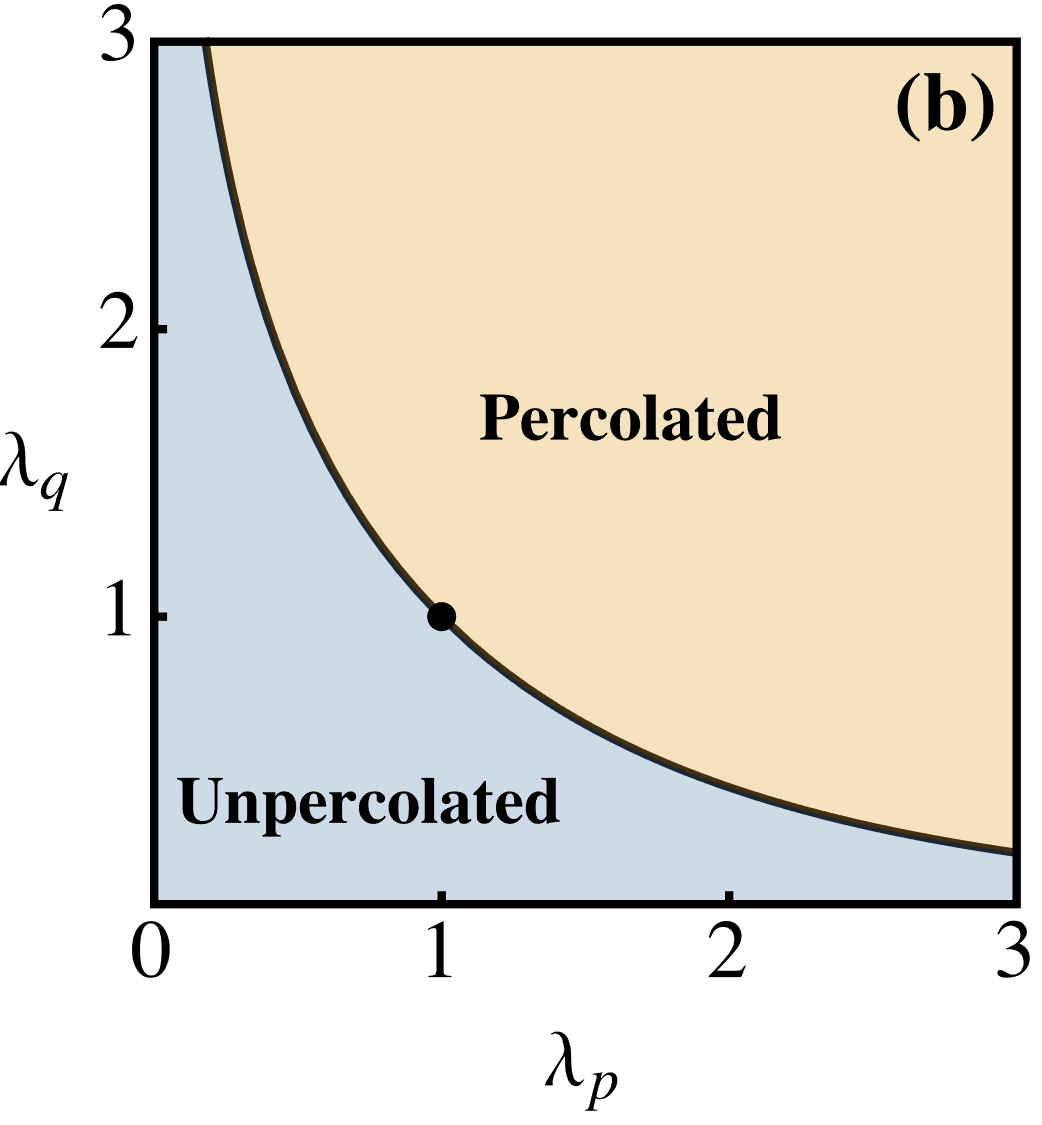}
    
    \caption{
    $(\lambda_p,\lambda_q)$ phase diagrams for \textbf{(a)} intransitive nonreciprocal random networks \textbf{(b)} transitive nonreciprocal random networks. The solid black lines in each plot correspond to the critical lines where the phase transitions occur and are given by Eqs.~(\ref{eq:percolation line intransitive}) and (\ref{eq:percolation line transitive}).
    }\label{fig:Phase plots}
\end{figure}

Fig.~\ref{fig:Phase plots} shows a side-by-side comparison of the phase-spaces of the intransitive and transitive networks as determined by Eqs.~(\ref{eq:percolation line intransitive}) and (\ref{eq:percolation line transitive}). In both plots, the critical line passes through the point $(\lambda_p,\lambda_q) = (1,1)$, yet a stark difference between them is that transitive networks contain a much larger region in which the network can be unpercolated. 

Moving on to the percolated state, we begin by noting that
\begin{multline}
    P(O\rightarrow\alpha) = \\ 
    1 -(1-p)^{(N-1)F_O(\alpha)}
    (1-q)^{(N-1)[O-F_O(\alpha)]}
\end{multline}
and
\begin{multline}
    P(\alpha\rightarrow I) = \\
    1 -(1-q)^{(N-1)F_I(\alpha)}
    (1-p)^{(N-1)[I-F_I(\alpha)]} \,,
\end{multline}
such that in the large $N$ limit
\begin{align}
    P(O\rightarrow\alpha) &= 
    1-e^{-\lambda_qO+(\lambda_q-\lambda_p)F_O} \,,\\
    P(\alpha\rightarrow I) &= 
    1-e^{-\lambda_pI-(\lambda_q-\lambda_p)F_I} \,.
\end{align}
Accordingly, Eqs.~(\ref{eq:ODE 1}), (\ref{eq:ODE 2}) and (\ref{eq:ODE 3}) can be written as
\begin{multline}
    F_S'(\alpha) = 
    \left[1-e^{-\lambda_qO+(\lambda_q-\lambda_p)F_O}\right]
    \times \\\times
    \left[1-e^{-\lambda_pI-(\lambda_q-\lambda_p)F_I}\right] \,,
    \label{eq:S ODE transitive}
\end{multline}
and
\begin{align}
    F_O'(\alpha) &= 
    1-e^{-\lambda_qO+(\lambda_q-\lambda_p)F_O} \,, \\
    F_I'(\alpha) &= 
    1-e^{-\lambda_pI-(\lambda_q-\lambda_p)F_I} \,.
\end{align}
These three coupled nonlinear ODEs determine the cumulative probability functions $F_S$, $F_O$ and $F_I$ and thus the percolation properties of the network. As in the intransitive case, the equations for $F'_O$ and $F'_I$ have decoupled and in fact have exact solutions
\begin{align}
    F_O(\alpha) &= 
    \frac{\lambda_qO-\ln
    \left[1\pm\left(e^{\lambda_qO}-1\right)
    e^{-(\lambda_q-\lambda_p)\alpha}\right]}
    {\lambda_q-\lambda_p} \,, \label{eq:transitive FO}\\
    F_I(\alpha) &= 
    -\frac{\lambda_pI-\ln
    \left[1\pm\left(e^{\lambda_pI}-1\right)
    e^{(\lambda_q-\lambda_p)\alpha}\right]}
    {\lambda_q-\lambda_p} \,,
    \label{eq:transitive FI}
\end{align}
where the $\pm$ correspond to two branches which each solve the equations. Subbing $\alpha=1$, we obtain implicit equations for $O$ and $I$
\begin{align}
    O &= 
    \frac{\lambda_qO-\ln
    \left[1\pm\left(e^{\lambda_qO}-1\right)
    e^{-(\lambda_q-\lambda_p)}\right]}
    {\lambda_q-\lambda_p} \,,\\
    I &= 
    -\frac{\lambda_pI-\ln
    \left[1\pm\left(e^{\lambda_pI}-1\right)
    e^{\lambda_q-\lambda_p}\right]}
    {\lambda_q-\lambda_p} \,.
\end{align}
or after some simplification
\begin{align}
    e^{\lambda_qO}-1 &= 
    \pm\left(e^{\lambda_pO}-1\right)e^{\lambda_q-\lambda_p}
     \label{eq:O implicit transitive}\,,\\
    e^{\lambda_qI}-1 &= 
    \pm\left(e^{\lambda_pI}-1\right)
    e^{\lambda_q-\lambda_p} 
    \label{eq:I implicit transitive}\,.
\end{align}
Notice that these are the same equation for $O$ and $I$ thus we have $O=I$, as we had in the intransitive network. Indeed, by the symmetry of the network model with respect to reversing the direction of all the links, this conclusion could have been determined from the start. More importantly, since $O\ge0$ and $I\ge0$, the left-hand side of these equations cannot be negative and thus the $+$ branch must contain the solution. In addition to the trivial solution $O=I=0$, there can exist another unique non-vanishing solution which is easy to determine numerically for any given $\lambda_p$ and $\lambda_q$.

Though an explicit form form for $O$ and $I$ is unavailable, near percolation, $O,I\ll 1$ such that we can expand these equations in $O$ and $I$ to obtain the leading order approximation
\begin{equation}
    O=I \approx 
    \begin{cases}
        0 & \lambda_q/\lambda_p < e^{\lambda_q-\lambda_p} \\
        2\frac{\lambda_q-\lambda_pe^{\lambda_q-\lambda_p}}
        {\lambda_p^2e^{\lambda_q-\lambda_p}-\lambda_q^2} &
        \lambda_q/\lambda_p > e^{\lambda_q-\lambda_p}
    \end{cases} \,.
    \label{eq:small O I expansion}
\end{equation}
It is easy to see from this expression that the percolation transition indeed occurs on exactly the line predicted by Eqs.~(\ref{eq:transitive unpercolated condition}) and (\ref{eq:transitive percolated condition}).

We can now return to Eq.~(\ref{eq:S ODE transitive}) which can be written explicitly as
\begin{multline}
    F_{S}'(\alpha) = 
    \frac{e^{\lambda_qO}-1}
    {1+\left(e^{\lambda_qO}-1\right)
    e^{-(\lambda_q-\lambda_p)\alpha}}
    \times \\ \times
    \frac{e^{\lambda_pI}-1}{1+\left(e^{\lambda_pI}-1\right)
    e^{(\lambda_q-\lambda_p)\alpha}} \,.
\end{multline}
Integrating over $\alpha$, we obtain
\begin{equation}
    F_{S}(\alpha) = 
    \frac{\left(e^{\lambda_{q}O}-1\right)
    \left(e^{\lambda_{p}I}-1\right)}
    {1-\left(e^{\lambda_{q}O}-1\right)
    \left(e^{\lambda_{p}I}-1\right)}
    [\alpha-F_{O}(\alpha)-F_{I}(\alpha)] \,,
    \label{eq:transitive FS}
\end{equation}
where $F_O(\alpha)$ and $F_I(\alpha)$ are given in Eqs.~(\ref{eq:transitive FO}) and (\ref{eq:transitive FI}). Substituting in $\alpha = 1$, we obtain
\begin{equation}
    S = \frac{(1-O-I) 
    \left(e^{\lambda_qO}-1\right)
    \left(e^{\lambda_pI}-1\right)}
    {1-\left(e^{\lambda_qO}-1\right)
    \left(e^{\lambda_pI}-1\right)} \,,
    \label{eq:S transitive}
\end{equation}
where we have used Eqs.~(\ref{eq:O implicit transitive}) and (\ref{eq:I implicit transitive}) to simplify the result. Near percolation, we can again expand in small $O$ and $I$ to write
\begin{equation}
    S \approx \lambda_p\lambda_qOI\,,
\end{equation}
or after subbing in Eq.~(\ref{eq:small O I expansion})
\begin{equation}
    S \approx 
    \begin{cases}
        0 & \lambda_q/\lambda_p < e^{\lambda_q-\lambda_p}\\
        \frac{\lambda_q}{\lambda_p}
        \left(2\frac{\lambda_q/\lambda_p-
        e^{\lambda_q-\lambda_p}}
        {e^{\lambda_q-\lambda_p}-(\lambda_q/\lambda_p)^2}\right)^2 &
        \lambda_q/\lambda_p > e^{\lambda_q-\lambda_p}
    \end{cases}
    \,.
\end{equation}
Unlike the intransitive network, here, $S\ne OI$. This is despite the fact that for each individual node, the out-degree and in-degree distributions are independent! This is of course because the joint distribution of a so-called ``typical node'', $\overline{P_\alpha(k_\mathrm{out},k_\mathrm{in})}$, given by Eq.~(\ref{eq:joint deg dist typical}), doesn't factor and as discussed in Sec.~\ref{sec:Transitive Degree Distribution}, this is an important characteristic feature of structured networks. In Appendix~\ref{sec:GF Method}, we compare the results obtained here with those using a naive application of the generating function method and show that the detailed network structure cannot simply be averaged into a so-called ``typical node''.

\section{\label{sec:Simulations}Simulations}

\subsection{Reciprocity}

To validate Eq.~(\ref{eq:mean reciprocity}), we generated $10^4$ intransitive networks and $10^4$ transitive networks, each consisting of $N=10^2$ nodes, for values $\lambda_p\in \{0,1/2,1,2,4\}$ and $\lambda_q\in [0,4]$. [Eqs.~(\ref{eq:lambda_p def}) and (\ref{eq:lambda_q def}) relate $\lambda_p$ and $\lambda_q$ to $p$ and $q$.] According to Eq.~(\ref{eq:mean reciprocity}), the mean reciprocity, $\left<\rho\right>$, tends to $0$ as $p,q\sim1/N\rightarrow 0 $ thus we have focused here on relatively small networks consisting of only $N=10^2$ nodes. 

\begin{figure}
    \centering
    \includegraphics[width=1.0\linewidth]{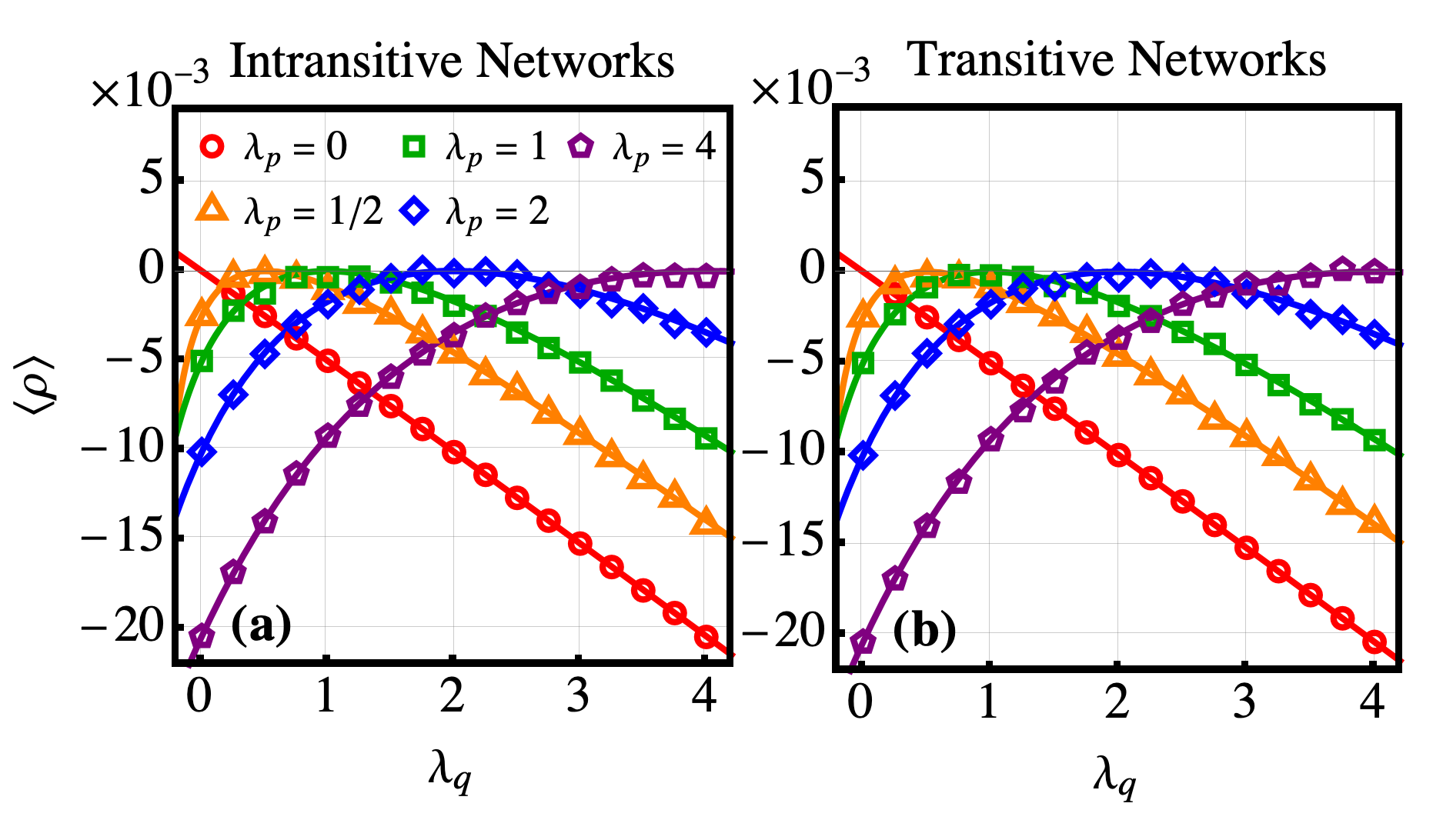}
    
    \caption{The mean reciprocity, $\left<\rho\right>$ as a function of $\lambda_q$ for $\lambda_p\in\{0,1/2,1,2,4\}$. The data points (symbols) are simulation data collected from $10^4$ \textbf{(a)} intransitive networks and \textbf{(b)} transitive networks, each consisting of $N=10^2$ nodes. The solid lines are given by Eq.~(\ref{eq:mean reciprocity}) and can be seen to predict the data superbly.}
    \label{fig:mean reciprocity}
\end{figure}

Fig.~\ref{fig:mean reciprocity} shows the mean reciprocity $\left<\rho\right>$ (symbols) as a function of $\lambda_q$. The solid lines in each plot are the theoretical value given by Eq.~(\ref{eq:mean reciprocity}). As can be seen, for all $\lambda_p$ and $\lambda_q$, Eq.~(\ref{eq:mean reciprocity}) performs superbly, demonstrating that nonreciprocal random networks are anti-reciprocal, as predicted. The derivation of $\left<\rho\right>$ in Appendix~\ref{sec:Reciprocity} doesn't depend on whether the network model is transitive or intransitive and this is demonstrated by the fact that Eq.~(\ref{eq:mean reciprocity}) superbly predicts the data points in both plots.

\subsection{Degree-Distribution Statistics}

To validate the predictions of Sec.~\ref{sec:Degree Distribution}, we generated $10^2$ intransitive networks and $10^2$ transitive networks, each consisting of $N = 10^4$ nodes for values $\lambda_p\in\{0,1/2,1,2,4\}$ and $\lambda_q\in[0,4]$. These networks were used to determine the mean out-degree and in-degree, the variance of the out-degree and in-degree distributions and the covariance of the joint degree distribution. For each of these measures, the calculation was performed in two ways: 
\begin{enumerate}
    \item Calculating the measure over all nodes in each network, followed by an averaging of the measure over all networks, i.e. $\left<\,\overline{\vphantom{X}\ldots}\,\right>$.
    \item For each node, calculating the measure over all networks, followed by averaging over all nodes, i.e. $\overline{\left<\ldots\right>}$. 
\end{enumerate}

\begin{figure}
    \centering
    \includegraphics[width=1.0\linewidth]{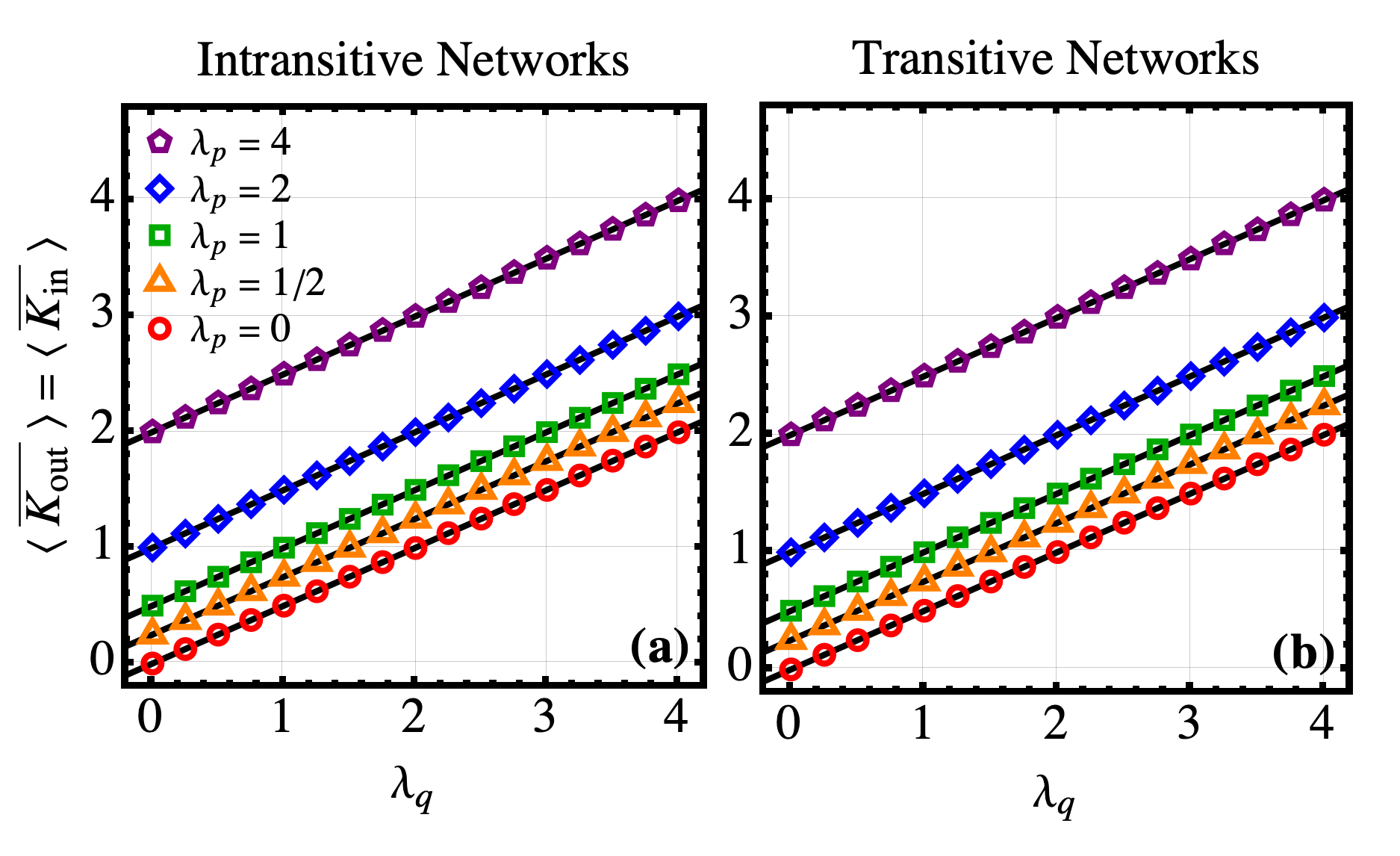}
    
    \caption{The mean out-degree $\left<\overline{K_\mathrm{out}}\right>$ (which is identical to the mean in-degree $\left<\overline{K_\mathrm{in}}\right>$) as a function of $\lambda_q$ for $\lambda_p \in\{0,1/2,1,2,4\}$. The data points (symbols) are collected by averaging over $10^2$ \textbf{(a)} intransitive networks and \textbf{(b)} transitive networks, each consisting of $N=10^4$ nodes. The solid lines are the theoretical values predicted by Eqs.~(\ref{eq:degree statistics intransitive}) and (\ref{eq:transitive mean degree}), for plots \textbf{(a)} and \textbf{(b)} respectively, and can be seen to superbly match the data.
    }
    \label{fig:mean degree}
\end{figure}

Fig.~\ref{fig:mean degree} shows the mean out-degree $\left<\overline{K_\mathrm{out}}\right>$ (which is necessarily identical to the mean in-degree $\left<\overline{K_\mathrm{in}}\right>$) for the intransitive and transitive networks. For this particular measure, the order of averaging can make no difference and thus we only present data calculated by first averaging over each network. The solid lines are the theoretical values predicted by Eqs.~(\ref{eq:degree statistics intransitive}) and (\ref{eq:transitive mean degree}) and can be seen to superbly predict the data. 

\begin{figure}
    \centering
    \includegraphics[width=1.0\linewidth]{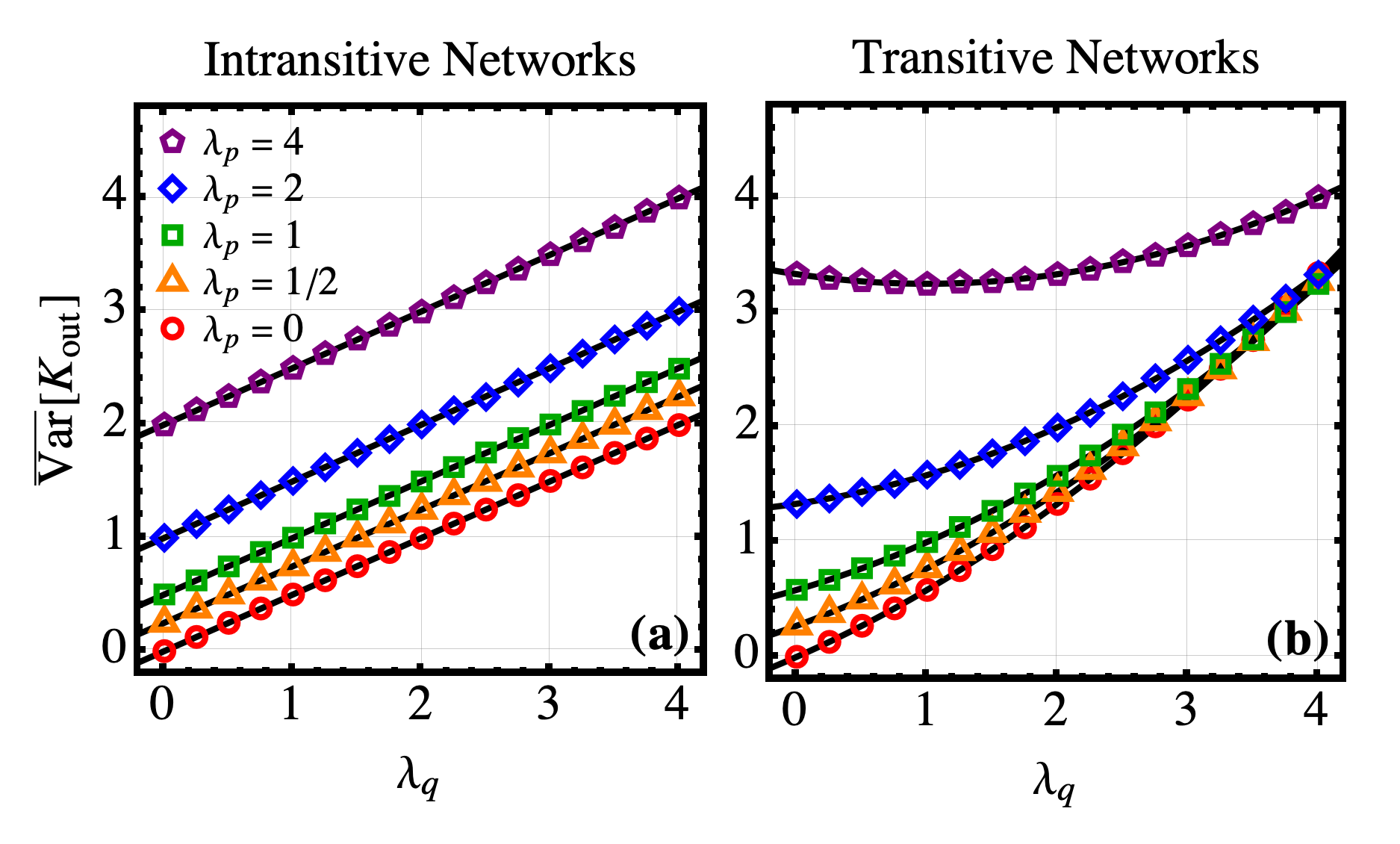}
    
    \vspace{-0.25 cm}
    
    \includegraphics[width=1.0\linewidth]{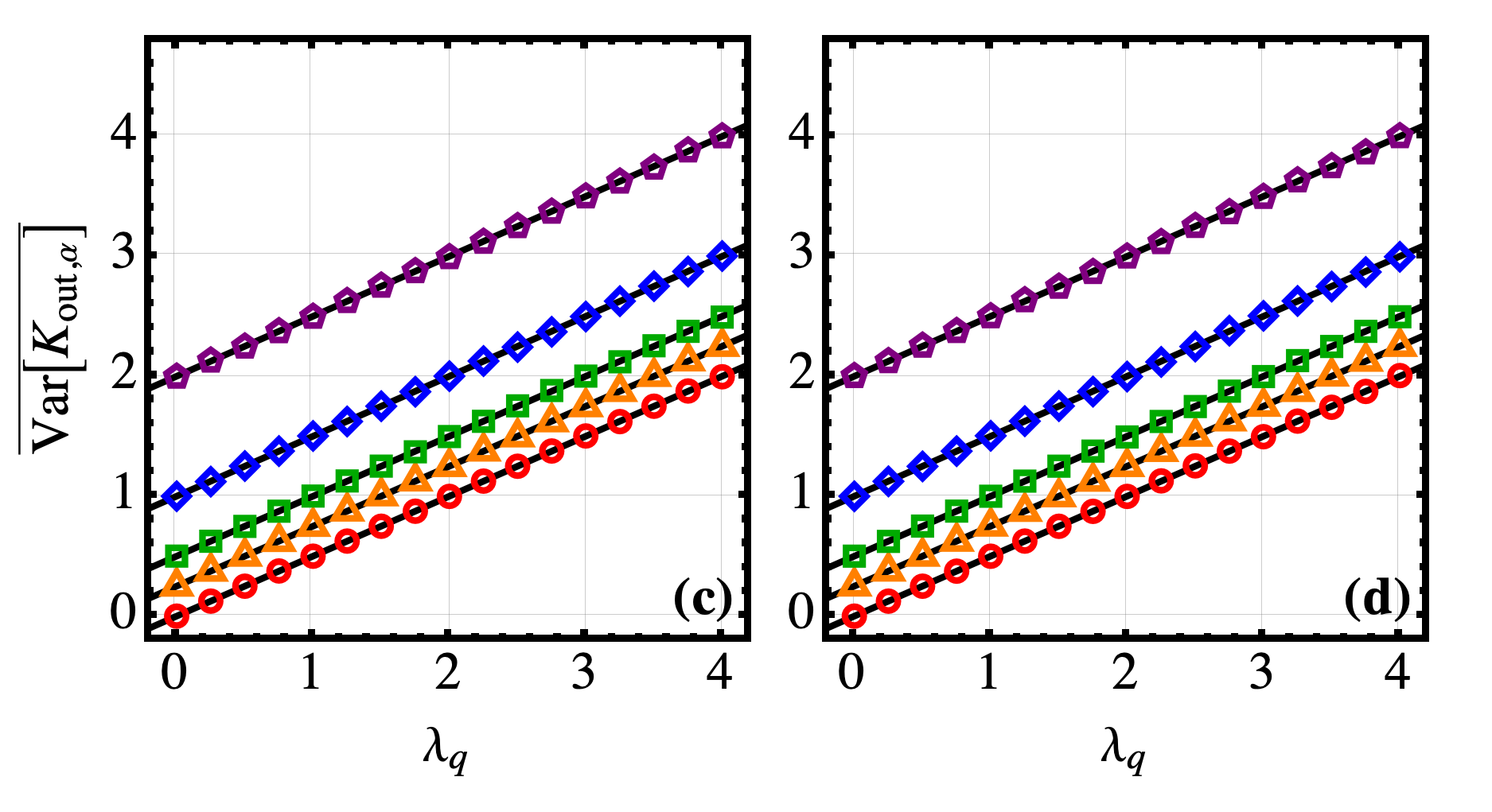}
    
    \caption{The variance of the out-degree distribution, calculated in two ways, as a function of $\lambda_q$ for $\lambda_p\in\{0,1/2,1,2,4\}$. Plots \textbf{(a)} and \textbf{(c)} correspond to simulation data collected from $10^2$ intransitive networks with $N=10^4$ nodes each. Plots \textbf{(b)} and \textbf{(d)} are the same but for transitive networks. The solid lines in plots \textbf{(a)} and \textbf{(c)} are the predicted values given by Eq.~(\ref{eq:degree statistics intransitive}). The solid lines in plot \textbf{(b)} are the values predicted by Eq.~(\ref{eq:transitive variance}) while those in plot \textbf{(d)} are obtained by integrating Eq.~(\ref{eq:transitive mean and var vs alpha}) over $\alpha$ from 0 to 1. In all cases, the theory perfectly matches the simulation results.
    }
    \label{fig:out-var degree}
\end{figure}

More interestingly, Fig.~\ref{fig:out-var degree} shows the variance of the out-degree distribution for intransitive and transitive networks calculated using the two methods described. It can be seen that for the intransitive networks [plots \ref{fig:out-var degree}(a) and \ref{fig:out-var degree}(c)], the order of averaging makes no difference. The solid lines are the values predicted by Eq.~(\ref{eq:degree statistics intransitive}) and can be seen to superbly predict the simulation results. In contrast, for the transitive networks [plots \ref{fig:out-var degree}(b) and \ref{fig:out-var degree}(d)], the ordering of averaging makes a large difference. The solid lines in plot \ref{fig:out-var degree}(b) are the values predicted by Eq.~(\ref{eq:transitive variance}) while those presented in plot \ref{fig:out-var degree}(d) are the values predicted by integrating Eq.~(\ref{eq:transitive mean and var vs alpha}) over $\alpha$ from 0 to 1. Again, the fit between data and theory is superb, validating our assertion that the order of averaging can matter greatly for structured networks but not for unstructured ones.

Because of the symmetry of our models with respect to swapping the direction of all links, the results for the variance of the in-degree distribution are practically identical to that of the out-degree distribution and thus we do not bother to present that data here.

\begin{figure}
    \centering
    \includegraphics[width=1.0\linewidth]{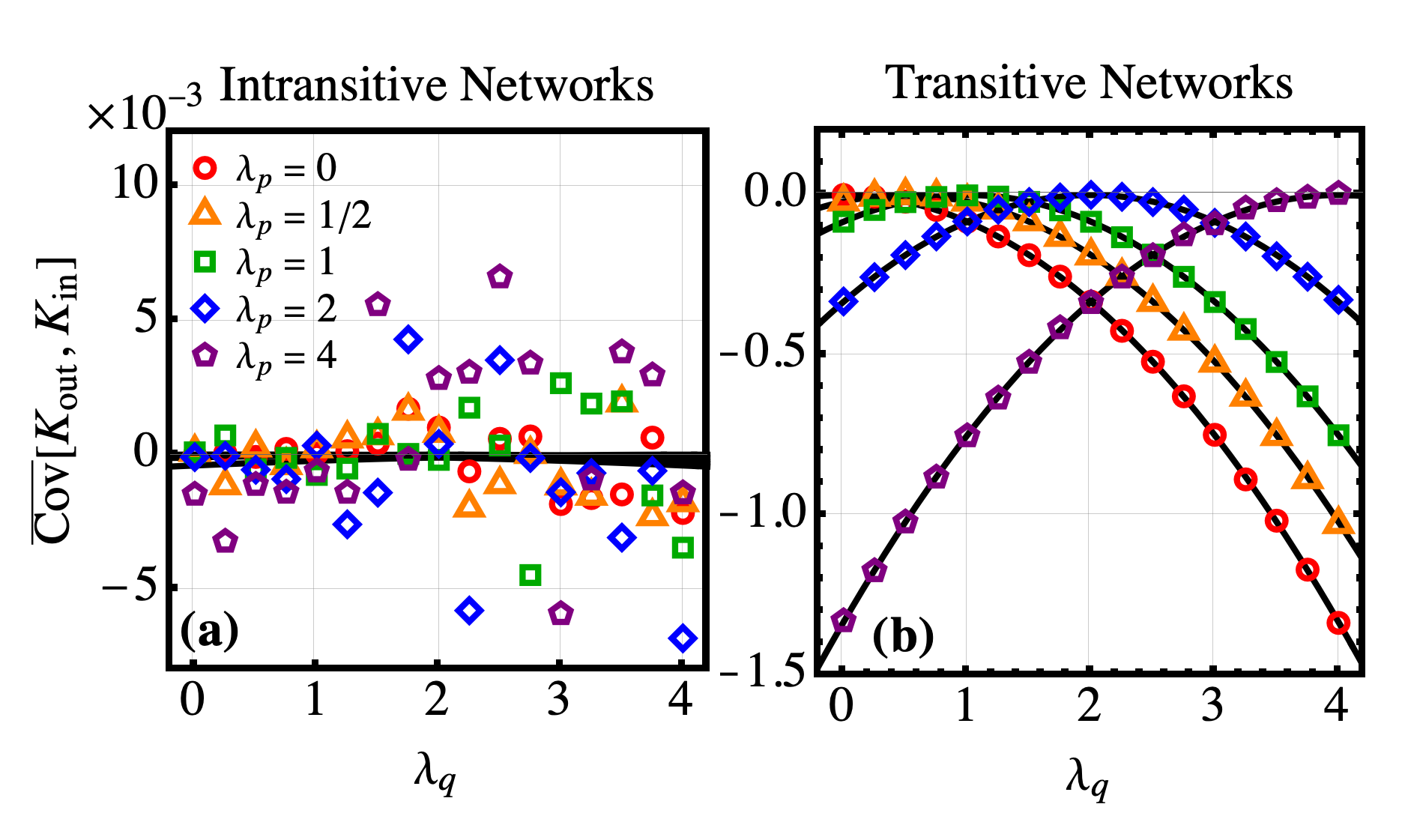}
    
    \vspace{-0.9cm}
    
    \includegraphics[width=1.0\linewidth]{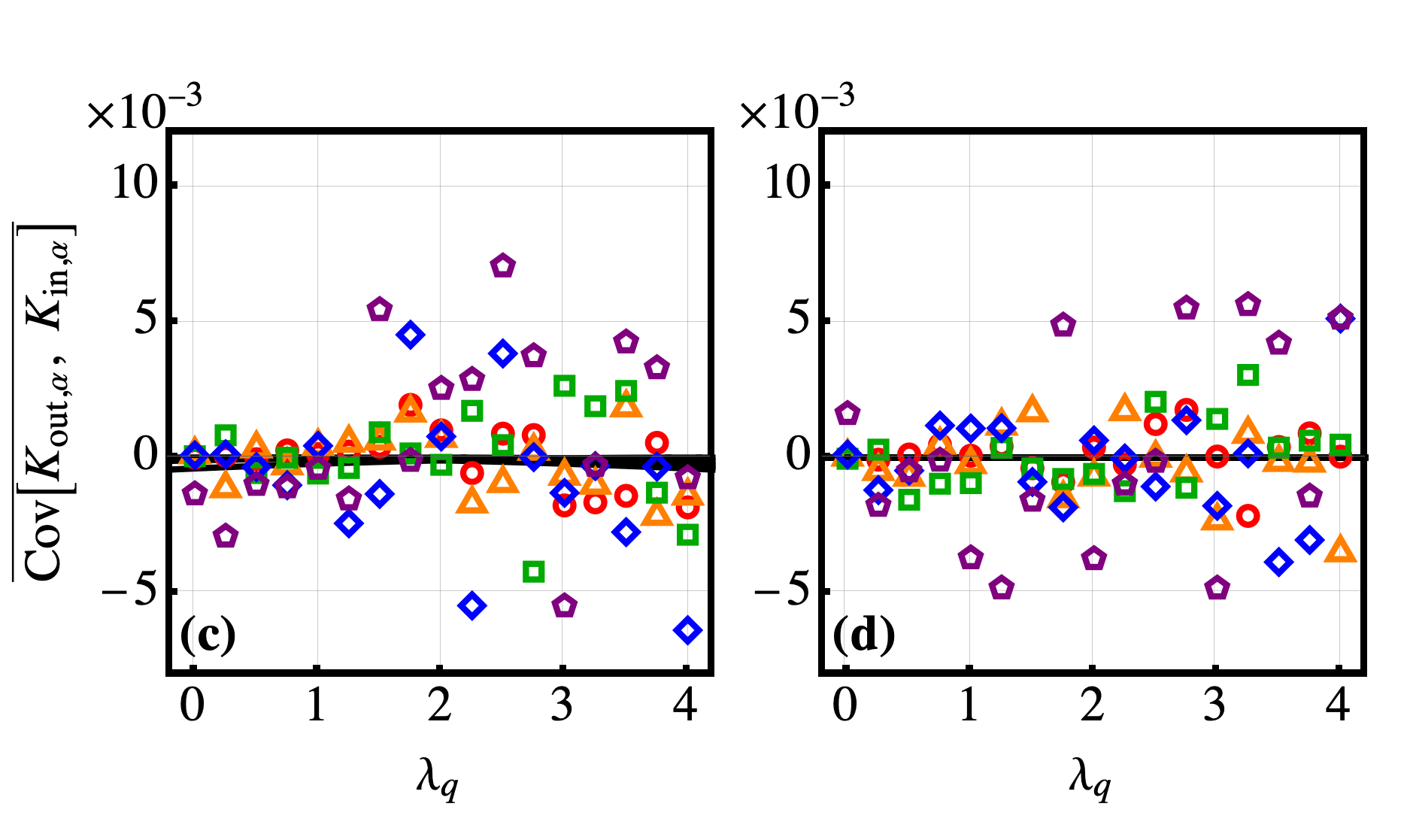}
    
    \caption{The covariance of the joint degree distribution, calculated in two ways, as a function of $\lambda_q$ for $\lambda_p\in\{0,1/2,1,2,4\}$. Plots \textbf{(a)} and \textbf{(c)} correspond to simulation data collected from $10^2$ intransitive networks with $N=10^4$ nodes each. Plots Plots \textbf{(b)} and \textbf{(d)} are the same but for transitive networks. The solid lines in plots \textbf{(a)} and \textbf{(c)} are the predicted values given by Eq.~(\ref{eq:intransitive covariance}). The solid lines in plot \textbf{(b)} which beautifully match the data are the values predicted by Eq.~(\ref{eq:transitive covariance}) while those in plot \textbf{(d)} are simply Eq.~(\ref{eq:transitive covariance 2}). In plots~\textbf{(a)}, \textbf{(b)} and \textbf{(d)}, the measured covariance is close to 0 and two or three orders of magnitude smaller than that shown in plot~\textbf{(b)} which accounts for the large amount of noise in this data.
    }
    \label{fig:cov degree}
\end{figure}

Finally, Fig.~\ref{fig:cov degree} shows the covariance of the joint degree distribution calculated in both ways for intransitive and transitive networks. Again, only for the transitive networks does the method of computation make a difference. The solid lines shown in plot~(b) are given by Eq.~(\ref{eq:transitive covariance}) and can be seen to beautifully match the data. In contrast, for plots~(a), (c) and (d), the predicted covariance, given by Eqs.~(\ref{eq:intransitive covariance}) and (\ref{eq:transitive covariance 2}), tends to 0 for large $N$ and thus $10^2$ networks is insufficient to overcome the large noise-to-signal ratio of this data. These plots never-the-less show that the measured covariance in these cases is two or three orders of magnitude smaller than that presented in plot~(b).

\subsection{Percolation\label{sec:Percolation Simuations}}

To validate the predictions of Sec.~\ref{sec:Percolation}, we used the sets of $10^2$ intransitive networks and $10^2$ transitive networks from the previous subsection, each consisting of $N=10^4$ nodes, to determine the size of the GSCC, the GOC and the GIC. Fig.~\ref{fig:S,O,I} shows the resulting data for $\lambda_p\in\{0,1/2,1,2,4\}$ as a function of $\lambda_q\in[0,4]$. The solid lines in plots~(a), (c) and (e) are the predicted values given by Eqs.~(\ref{eq:O, I intransitive}) and (\ref{eq:S intransitive}). The solid lines in plots~(b), (d) and (f) are derived from Eqs.~(\ref{eq:O implicit transitive}), (\ref{eq:I implicit transitive}) and (\ref{eq:S transitive}). As can be seen, in all cases, the theory beautifully predicts the simulation results.

\begin{figure}
    \centering
    \includegraphics[width=1.0\linewidth]{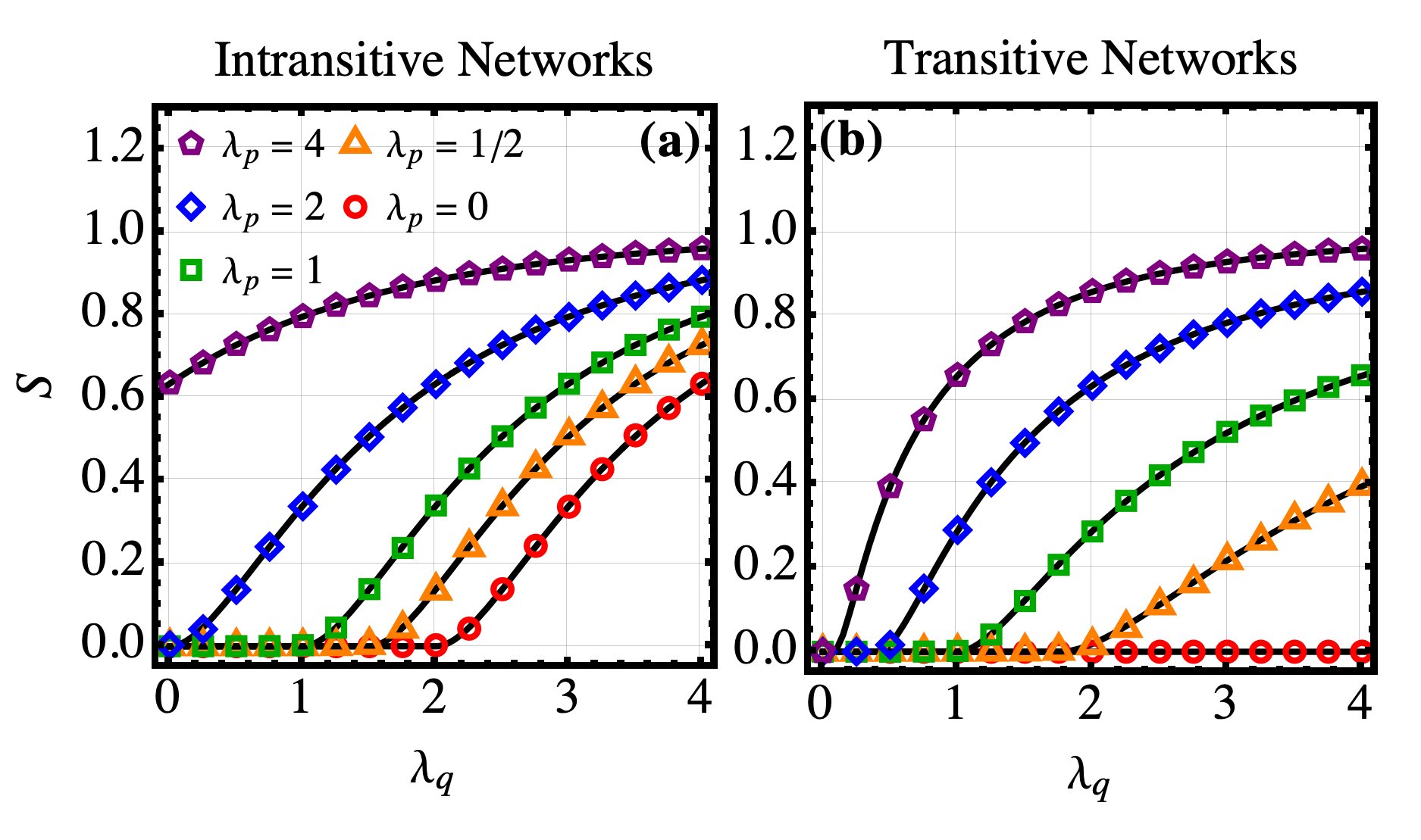}
    
    \vspace{-0.4cm}
    
    \includegraphics[width=1.0\linewidth]{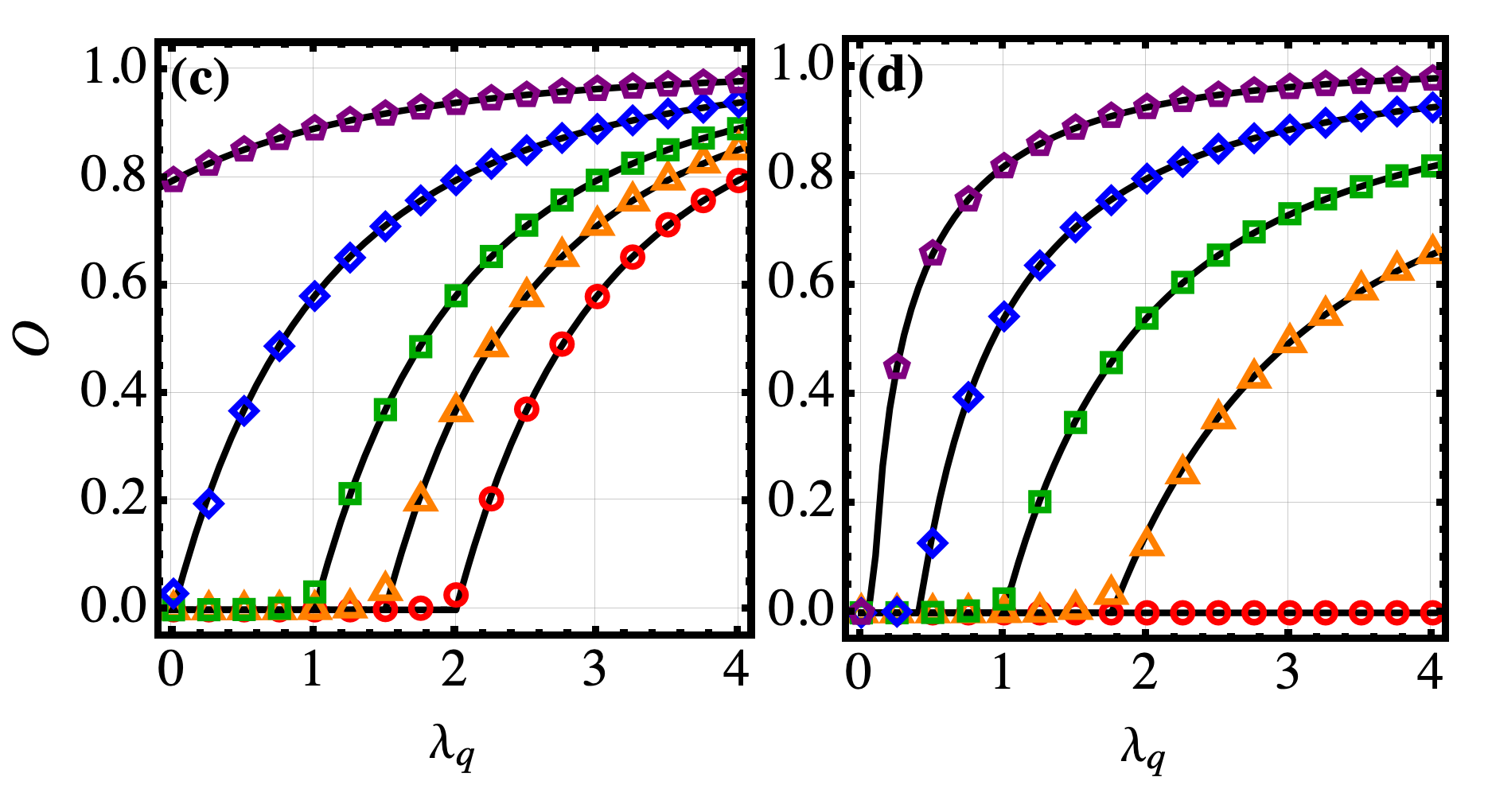}

    \vspace{-0.4cm}
    
    \includegraphics[width=1.0\linewidth]{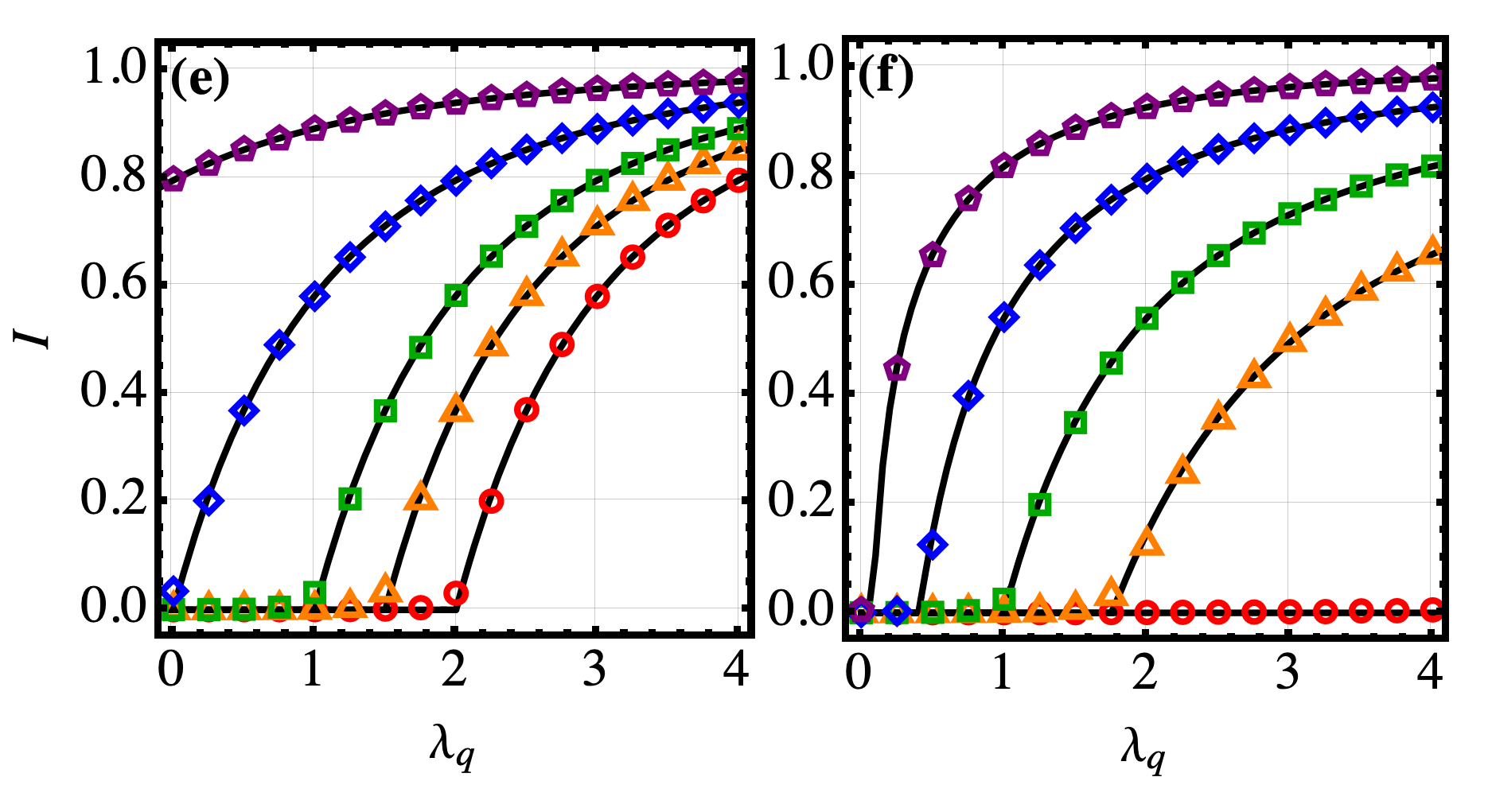}
    
    \caption{The size of the GSCC, $S$, the GOC, $O$, and the GIC, $I$, for intransitive and transitive networks, as a function of $\lambda_q$ for $\lambda_p\in\{0,1/2,1,2,4\}$. Each data point is determined by averaging $10^2$ networks, each containing $N=10^4$ nodes. The solid lines in plots (a), (c) and (e) are the predicted values given by Eqs.~(\ref{eq:O, I intransitive}) and (\ref{eq:S intransitive}). The solid lines in plots~(b), (d) and (f) are derived from Eqs.~(\ref{eq:O implicit transitive}), (\ref{eq:I implicit transitive}) and (\ref{eq:S transitive}). In every plot, the theory beautifully predicts the simulations.
    }
    \label{fig:S,O,I}
\end{figure}

While the giant components of the intransitive and transitive networks behave similarly, there are important differences, especially when either $\lambda_p$ is large and $\lambda_q$ is small or vice versa. For instance, when $\lambda_p=4$, the intransitive networks are percolated for all values of $\lambda_q$ while the transitive networks are unpercolated for sufficiently small values of $\lambda_q$. Similarly, when $\lambda_p=0$, the intransitive networks percolate for sufficiently large $\lambda_q$ while the transitive networks never percolate. These results are consistent with those predicted in Fig.~\ref{fig:Phase plots} and emphasise the important role structure plays in determining the percolation properties of nonreciprocal random networks.

In addition to the sizes of the GSCC, the GOC and the GIC, the theory developed in Sec.~\ref{sec:Percolation} predicts the cumulative probability functions $F_S(\alpha)$, $F_O(\alpha)$ and $F_I(\alpha)$ that node $\alpha=(i-1)/(N-1)$ will belong to the respective giant cluster. While these functions are trivially linear for unstructured networks, structured networks are characterised by having non-trivial cumulative probability functions.

\begin{figure}
    \centering
    \includegraphics[width=1.0\linewidth]{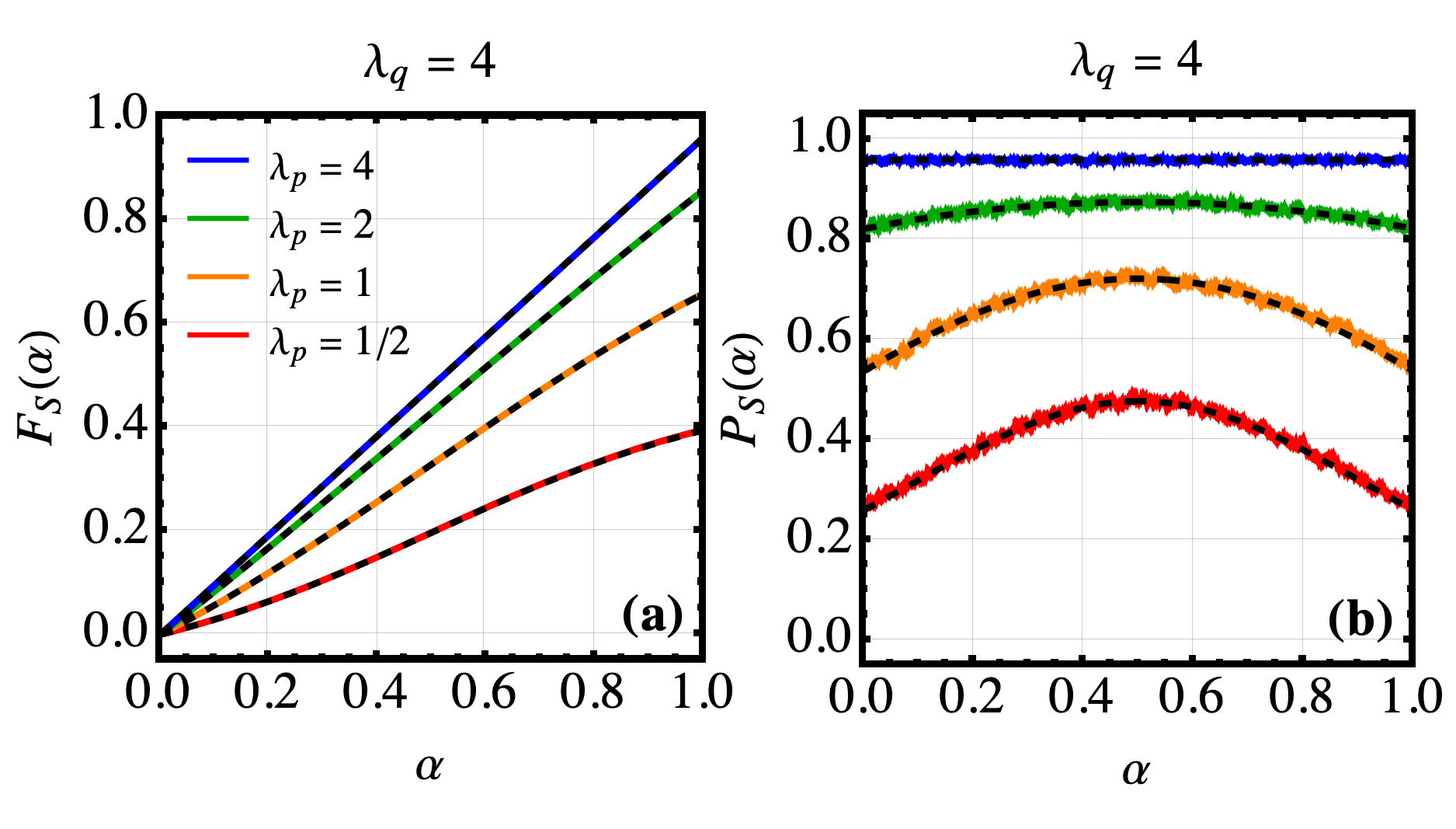}
    
    \vspace{-0.4cm}
    
    \includegraphics[width=1.0\linewidth]{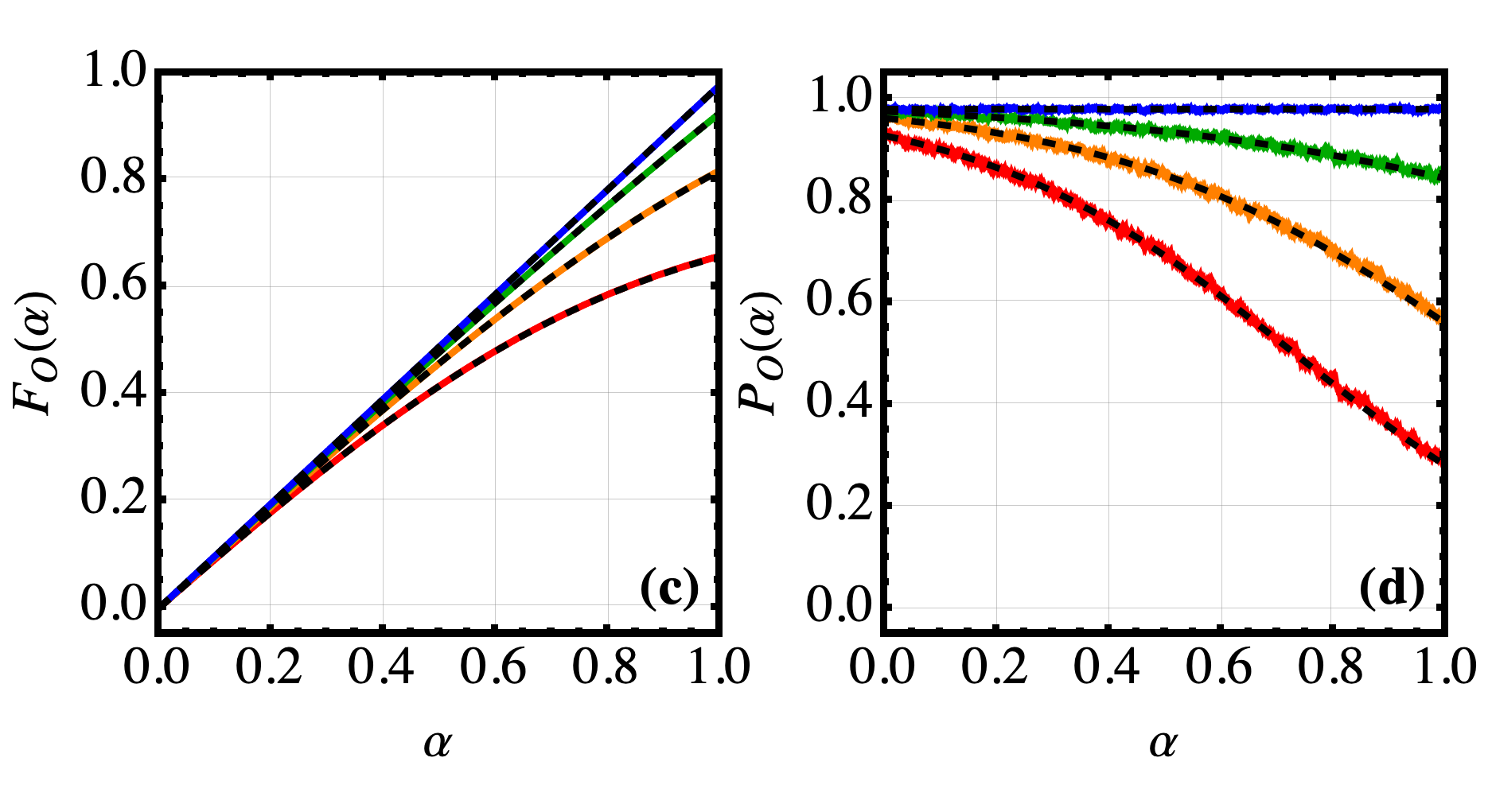}

    \vspace{-0.4cm}
    
    \includegraphics[width=1.0\linewidth]{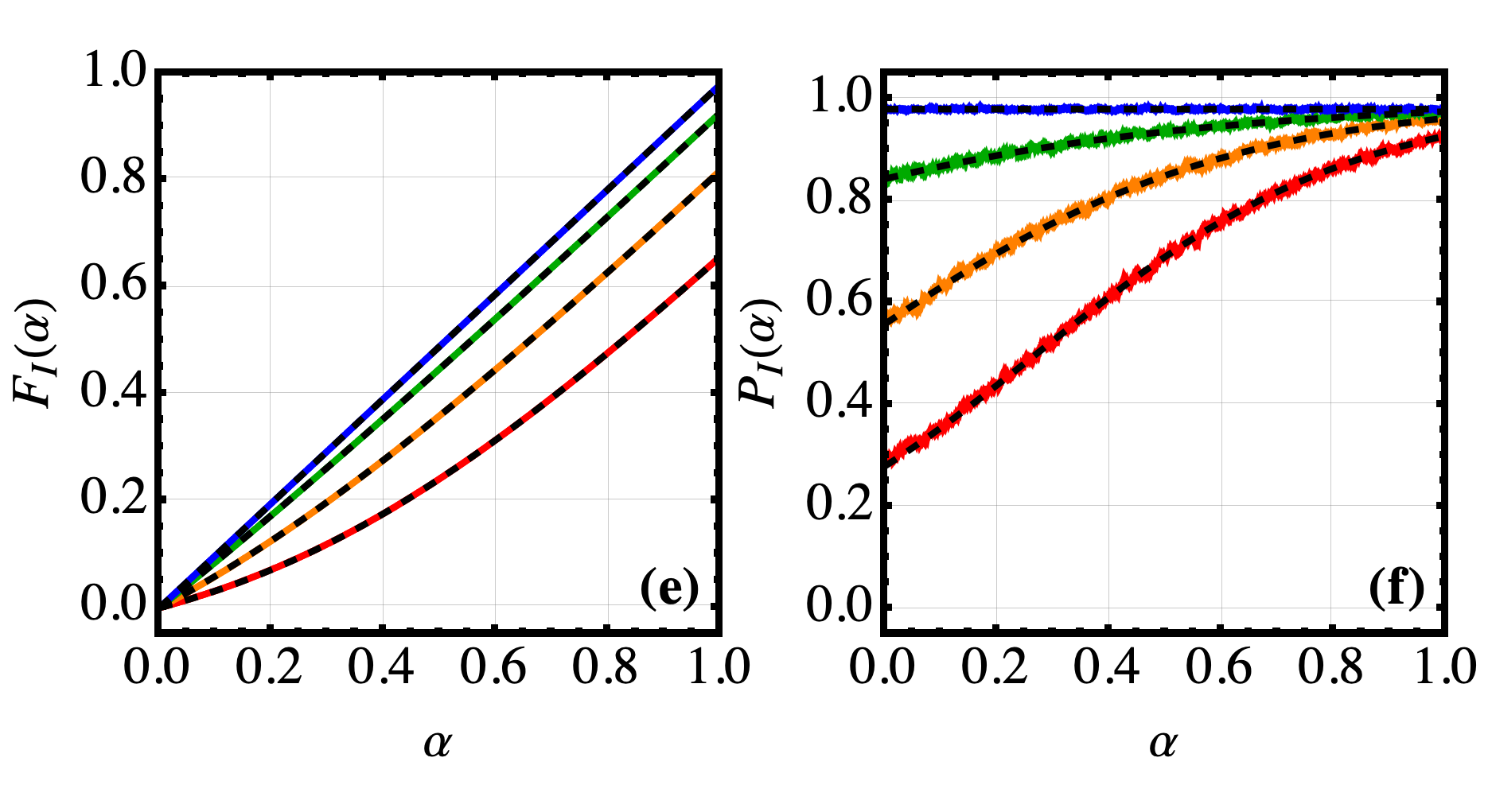}
    
    \caption{The cumulative probability functions, $F_S(\alpha)$, $F_O(\alpha)$ and $F_I(\alpha)$ and their corresponding probability distributions $P_S(\alpha)=F'_S(\alpha)$, $P_O(\alpha)=F'_O(\alpha)$ and $P_I(\alpha)=F'_I(\alpha)$ as functions of $\alpha$, for transitive networks with values $\lambda_q = 4$ and $\lambda_p\in\{1/2,1,2,4\}$. Each data set was obtained by averaging over $10^2$ networks, each consisting of $N=10^4$ nodes. The black dashed lines are the theory predicted by Eqs.~(\ref{eq:transitive FO}), (\ref{eq:transitive FI}) and (\ref{eq:transitive FS}) and can be seen to match the data superbly.
    }
    \label{fig:FS,FO,FI}
\end{figure}

Fig.~\ref{fig:FS,FO,FI} shows these cumulative probability functions as well as their respective probability distributions $P_S(\alpha)=F_S'(\alpha)$, $P_O(\alpha)=F_O'(\alpha)$ and $P_I(\alpha)=F_I'(\alpha)$, for transitive networks with the particular values $\lambda_q=4$ and $\lambda_p\in\{1/2,1,2,4\}$. The data for the probability distribution functions were derived from the cumulative probability functions by numerically differentiating each data set with respect to $\alpha$. This however results in highly noisy data thus a running average with a window consisting of 100 data points was applied to smooth the data. Even with this smoothing, the data presented in plots~(b), (d) and (f) are substantially noisier than that presented in plots~(a), (c) and (e). The dashed black lines are the theoretical values predicted by Eqs.~(\ref{eq:transitive FO}), (\ref{eq:transitive FI}) and (\ref{eq:transitive FS}) and their derivatives, and can be seen to match the data superbly.

As predicted, the cumulative probability functions for transitive structured networks are not linear and their corresponding probability distributions are thus not uniform. The exception to this is when $\lambda_p=\lambda_q$. This is of course unsurprising as in the limit $\lambda_p\rightarrow\lambda_q$, transitive networks become increasingly indistinguishable from intransitive networks. Accordingly, not only has our theory correctly provided the sizes of the GSCC, the GOC and the GIC, it has correctly provided the probability that any particular node will belong to one of these giant clusters.

\section{\label{sec:Discussion}Discussion}

In this paper, we have studied nonreciprocal random networks in which the probability, $P_{i\rightarrow j}$, of node $i$ linking to node $j$ differs from the probability, $P_{j\rightarrow i}$, of node $j$ linking to node $i$. This model has allowed us to study the effect of network structure on networks in general and percolation in particular.
Of particular note, our results demonstrate how structure leads to the break down of commutativity of network averaging and ensemble averaging. Further, we have shown that structure can substantially modify the percolation properties of a network, in our case, by suppressing the percolated phase.

There are many further generalisations which could be studied to further illuminate the effects of network structure. For instance, as mentioned in Sec.~\ref{sec:Nonreciprocal Random Networks}, the transitivity of nonreciprocal networks can be continuously varied with a parameter $t$. As $t$ is increased from 0 (complete intransitivity) to 1 (complete transitivity), the percolation properties must vary between the results found in this paper. It would be interesting to consider how this transition occurs, for instance, does this transition occur smoothly or is there another phase transition hidden here?

Additionally, there are other possible structures which can be considered. The transitive structure is imposed by using the probabilities $p$ and $q$ to define a (partial) ordering and this essentially induces an underlying one-dimensional lattice structure on the network. An alternative might be to use $p$ and $q$ to induce a chiral periodic structure corresponding to a one-dimensional lattice wrapped into a ring shape. Other more complicated structures can also be envisioned and imposed with the probabilities $p$ and $q$.

Beyond these structures, even more complicated structures can be created by introducing additional probabilities beyond just $p$ and $q$. Indeed, though difficult, a systematic study of the network probability matrix formalism developed in \cite{Bollobas2007, McCulloh2007} seems like an important neglected project.

The technique developed in this paper to study percolation in structured networks also seems like a powerful tool with applications beyond the relatively straight-forward networks studied here. Reducing the percolation problem to the solution of integral and differential equations makes it approachable with well-known techniques from calculus, a valuable boon to the study of percolation and random networks. While it is unlikely that exact solutions will be attainable for most networks, it is clear that this technique has the potential to increase the number of exactly solvable models. Further, even when exact solutions are unattainable, the technique is capable of providing additional perspectives on percolation through numerical solutions.

\begin{acknowledgments}

The author would like to thank Peter Grassberger for directing their attention to the relevant literature on resistor-diode networks and biased directed percolation.

\end{acknowledgments}

\appendix
\section{\label{sec:Reciprocity}Reciprocity}

As discussed in Sec.~\ref{sec:Nonreciprocal Random Networks}, the reciprocity $\rho$ of a network with adjacency matrix $A$ is given by
\begin{equation}
    \rho[A] = \frac{r - \bar{A}}{1 - \bar{A}}\,,
    \label{eq:rho[A]}
\end{equation}
where $\bar{A} = \sum_{i\ne j}A_{ij}/N(N-1)$
denotes the link density and $r = L^\leftrightarrow/L$ denotes the ratio of the number of links $L^\leftrightarrow = \sum_{i\ne j}A_{ij}A_{ji}$ between reciprocal pairs of nodes and the total number of (non-self) links $L = \sum_{i\ne j}A_{ij}$.

For an ensemble of random networks, the reciprocity becomes a random variable whose mean is given by
\begin{equation}
    \left<\rho\right> = \sum_A \rho[A]P(A) \,,
\end{equation}
where the sum is taken over all possible adjacency matrices and $P(A)$ denotes the probability that a network with adjacency matrix $A$ will be instantiated. 

Let us consider nonreciprocal random networks with $N$ nodes and exactly two probabilities $p$ and $q$. Suppose that we instantiate a network with $n_p$ links that were created with probability $p$, $n_q$ links that were created with probability $q$, and $m$ pairs of reciprocal nodes. Then Eq.~(\ref{eq:rho[A]}) implies
\begin{equation}
    \rho(n_p,n_q,m) =
    \frac{\frac{2m}{n_p+n_q}-\frac{n_p+n_q}{2M}}
    {1-\frac{n_p+n_q}{2M}} \,,
\end{equation}
where we have denoted $M = N(N-1)/2$.
The probability of instantiating such a network is just
\begin{equation}
    P(n_p,n_q,m) = 
    p^{n_p}(1-p)^{M-n_p}q^{n_q}(1-q)^{M-n_q} \,.
\end{equation}
Accordingly, we can write the mean reciprocity as\begin{equation}
    \left<\rho\right> = \sum_{n_p,n_q,m} \Omega(n_p,n_q,m)\rho(n_p,n_q,m)P(n_p,n_q,m) \,,
    \label{eq:mean rho}
\end{equation}
where
\begin{multline}
    \Omega(n_p,n_q,m) = 
    \binom{M}{n_p}\binom{n_p}{m}\binom{M-n_p}{n_q-m} \\
    = \frac{M!}{(M-n_p-n_q+m)!(n_p-m)!(n_q-m)!m!}
\end{multline}
denotes the multiplicity of networks with values $n_p$, $n_q$ and $m$.

Upon substitution into Eq.~(\ref{eq:mean rho}), we obtain
\begin{multline}
    \left<\rho\right> =
    M!\sum_{n_p,n_q} 
    \frac{p^{n_p}(1-p)^{M-n_p}q^{n_q}(1-q)^{M-n_q}}
    {1-\frac{n_p+n_q}{2M}} \times\\
    \sum_m 
    \frac{\frac{2m}{n_p+n_q}-\frac{n_p+n_q}{2M}}{(M-n_p-n_q+m)!(n_p-m)!(n_q-m)!m!} \,.
\end{multline}
The sum over $m$ can be carried out with the aid of \textit{Mathematica} to give
\begin{multline}
    \sum_m 
    \frac{\frac{2m}{n_p+n_q}-\frac{n_p+n_q}{2M}}{(M-n_p-n_q+m)!(n_p-m)!(n_q-m)!m!} = \\
    2\frac{\,_{2}F_{1}(1-n_{p},1-n_{q};M-n_{p}-n_{q}+2;1)}{(n_{p}+n_{q})(n_{p}-1)!(n_{q}-1)!(M-n_{p}-n_{q}+1)!} + \\
    -\frac{(n_{p}+n_{q})\,_{2}F_{1}(-n_{p},-n_{q};M-n_{p}-n_{q}+1;1)}{2Mn_{p}!n_{q}!(M-n_{p}-n_{q})!}
\end{multline}
where $\,_{2}F_{1}(a,b;c;z)$ denotes the hypergeometric function. Using the Chu–Vandermonde identity \cite{NIST:DLMF}
\begin{equation}
    \,_{2}F_{1}(-k,b;c;1) = 
    \frac{\Gamma(c-b+k)\Gamma(c)}
    {\Gamma(c+k)\Gamma(c-b)} \qquad 
    k\in\mathbb{N} \,,
\end{equation}
where $\Gamma(z)$ denotes the Gamma function, this expression can be simplified to give
\begin{multline}
    \sum_m 
    \frac{\frac{2m}{n_p+n_q}-\frac{n_p+n_q}{2M}}{(M-n_p-n_q+m)!(n_p-m)!(n_q-m)!m!} = \\
    -\left(\frac{n_{p}+n_{q}}{2}-\frac{2n_{p}n_{q}}{n_{p}+n_{q}}\right)\frac{(M-1)!}{n_{p}!(M-n_{p})!n_{q}!(M-n_{q})!} \,.
\end{multline}
Accordingly, we have
\begin{multline}
    \left<\rho\right> =
    -\frac{1}{M} \sum_{n_p,n_q}
    \frac{\frac{n_p+n_q}{2}-\frac{2n_pn_q}{n_p+n_q}}
    {1-\frac{n_p+n_q}{2M}} \times \\ 
    \binom{M}{n_p}p^{n_p}(1-p)^{M-n_p}
    \binom{M}{n_q}q^{n_q}(1-q)^{M-n_q} \,.
\end{multline}
This sum is tricky to compute but easy to approximate. Since for even moderately large $M$, the second line of the summand is sharply peaked around $(n_p,n_q) \approx (Mp,Mq)$, we can approximate this as \cite{Kim2002, Lambiotte2016}
\begin{equation}
    \left<\rho\right>\approx
    -\frac{1}{M}
    \left[\frac{\frac{n_p+n_q}{2}-\frac{2n_pn_q}{n_p+n_q}}
    {1-\frac{n_p+n_q}{2M}}\right]_{(n_p,n_q)=(Mp,Mq)} \,,
\end{equation}
i.e.
\begin{equation}
    \left<\rho\right>\approx
    -\frac{\frac{p+q}{2}-\frac{2pq}{p+q}}
    {1-\frac{p+q}{2}}
    = -\frac{(p-q)^{2}}{(p+q)(2-p-q)}\,.
\end{equation}
Remarkably, the result doesn't depend on \mbox{$M = N(N-1)/2$} and thus doesn't depend on the number of nodes $N$. Since $\left<\rho\right> \le 0$, we find that nonreciprocal random networks are ``anti-reciprocal'', i.e. they are less likely to have reciprocal pairs of nodes than an ordinary  random network.

\section{\label{sec:GF Method}The Generating Function Method for Percolation}

The most common and arguably simplest framework for determining percolation is the generating function method \cite{Newman2001, Dorogovtsev2001}. For directed networks, this entails determining the size of the GSCC, GOC and GIC from the generating function $G(x,y)$ of the joint degree-distribution $P(k_\mathrm{out},k_\mathrm{in})$, [defined in Eq.~(\ref{eq:G(x,y) def})]. Given the generating function $G(x,y)$ we can define the two auxiliary functions 
\begin{align}
    G_1^{(O)}(x) &= 
    \frac{1}{\left\langle K_{\mathrm{out}}\right\rangle }
    \left.\frac{\partial G}{\partial y}\right|_{y=1} \,,\\
    G_1^{(I)}(y) &= 
    \frac{1}{\left\langle K_{\mathrm{in}}\right\rangle }
    \left.\frac{\partial G}{\partial x}\right|_{x=1} \,,
\end{align}
and the two values $x_0$ and $y_0$ as the solutions to the equations
\begin{align}
    x_0&=G_{1}^{(O)}(x_0) \label{eq:x0}\,,\\
    y_0&=G_{1}^{(I)}(y_0) \label{eq:y0}\,.
\end{align}
Then the sizes of $O$, $I$ and $S$ are given by
\begin{align}
    O&=1-G(x_0,1) \,, \label{eq:GF Method O}\\
    I&=1-G(1,y_0) \,,\label{eq:GF Method I}\\
    S &= G(x_0,y_0)+O+I-1 \,.
    \label{eq:GF Method S}
\end{align}

Here, we will briefly show that while the generating function method produces identical results for unstructured networks to the method developed in Sec.~\ref{sec:Percolation}, naive attempts to apply it to structured networks can result in unacceptable deviations.

\subsection{Intransitive Networks}

For the intransitive case, the generating function of the joint degree-distribution is given by Eq.~(\ref{eq:G(x,y) intransitive}) and thus
\begin{align}
    G_{1}^{(O)}(x)&=
    e^{\frac{\lambda_{p}+\lambda_{q}}{2}(x-1)} \,,\\
    G_{1}^{(I)}(y)&=
    e^{\frac{\lambda_{p}+\lambda_{q}}{2}(y-1)} \,.
\end{align}
Accordingly, the solutions to Eqs.~(\ref{eq:x0}) and (\ref{eq:y0}) are just
\begin{equation}
    x_{0} = y_{0} =
    -\frac{2}{\lambda_{p}+\lambda_{q}}
    W\left(-\frac{\lambda_{p}+\lambda_{q}}{2}
    e^{-\frac{\lambda_{p}+\lambda_{q}}{2}}\right) \,,
\end{equation}
where $W(z)$ denotes the principle branch of the Lambert-W function \cite{NIST:DLMF}. Upon substitution into Eqs.~(\ref{eq:GF Method O}), (\ref{eq:GF Method I}) and (\ref{eq:GF Method S}) we obtain precisely the same expressions as Eqs.~(\ref{eq:O, I intransitive}) and (\ref{eq:S intransitive}). Accordingly, the method developed in the text and the generating function produce exactly the same predictions for unstructured networks.

\subsection{Transitive Networks}

In Sec.~\ref{sec:Transitive Degree Distribution}, we showed that structured networks such as the transitive nonreciprocal random network do not have a unique joint degree-distribution but rather each node in the network has its own joint degree-distribution. Accordingly, if we wish to apply the generating function method to such networks, we need to consolidate these distributions into a single distribution with the most natural method being that of averaging them. Eq.~(\ref{eq:G(x,y) transitive}) gives the ``average'' generating function $\overline{G(x,y)}$ of the transitive nonreciprocal random network.

We can now determine
\begin{align}
    \overline{G_{1}^{(O)}(x)} &=
    \frac{[1+\lambda_{q}(x-1)]e^{\lambda_{p}(x-1)}-
    [1+\lambda_{p}(x-1)]e^{\lambda_{q}(x-1)}}
    {\frac{\lambda_{p}+\lambda_{q}}{2}
    (\lambda_{p}-\lambda_{q})(x-1)^{2}} \,,\\
    \overline{G_{1}^{(I)}(y)} &=
    \frac{[1+\lambda_{q}(y-1)]e^{\lambda_{p}(y-1)}-
    [1+\lambda_{p}(y-1)]e^{\lambda_{q}(y-1)}}
    {\frac{\lambda_{p}+\lambda_{q}}{2}
    (\lambda_{p}-\lambda_{q})(y-1)^{2}} \,,
\end{align}
which can be used to determine $x_0$ and $y_0$ through Eqs.~(\ref{eq:x0}) and Eqs.~(\ref{eq:y0}). Though we cannot write explicit equations for $x_0$ and $y_0$, it is easy enough to determine them numerically. Expressions for $O$, $I$ and $S$ can then be easily obtained from Eqs.~(\ref{eq:GF Method O}), (\ref{eq:GF Method I}) and (\ref{eq:GF Method S}).

\begin{figure}
    \centering
    \includegraphics[width=0.8\linewidth]{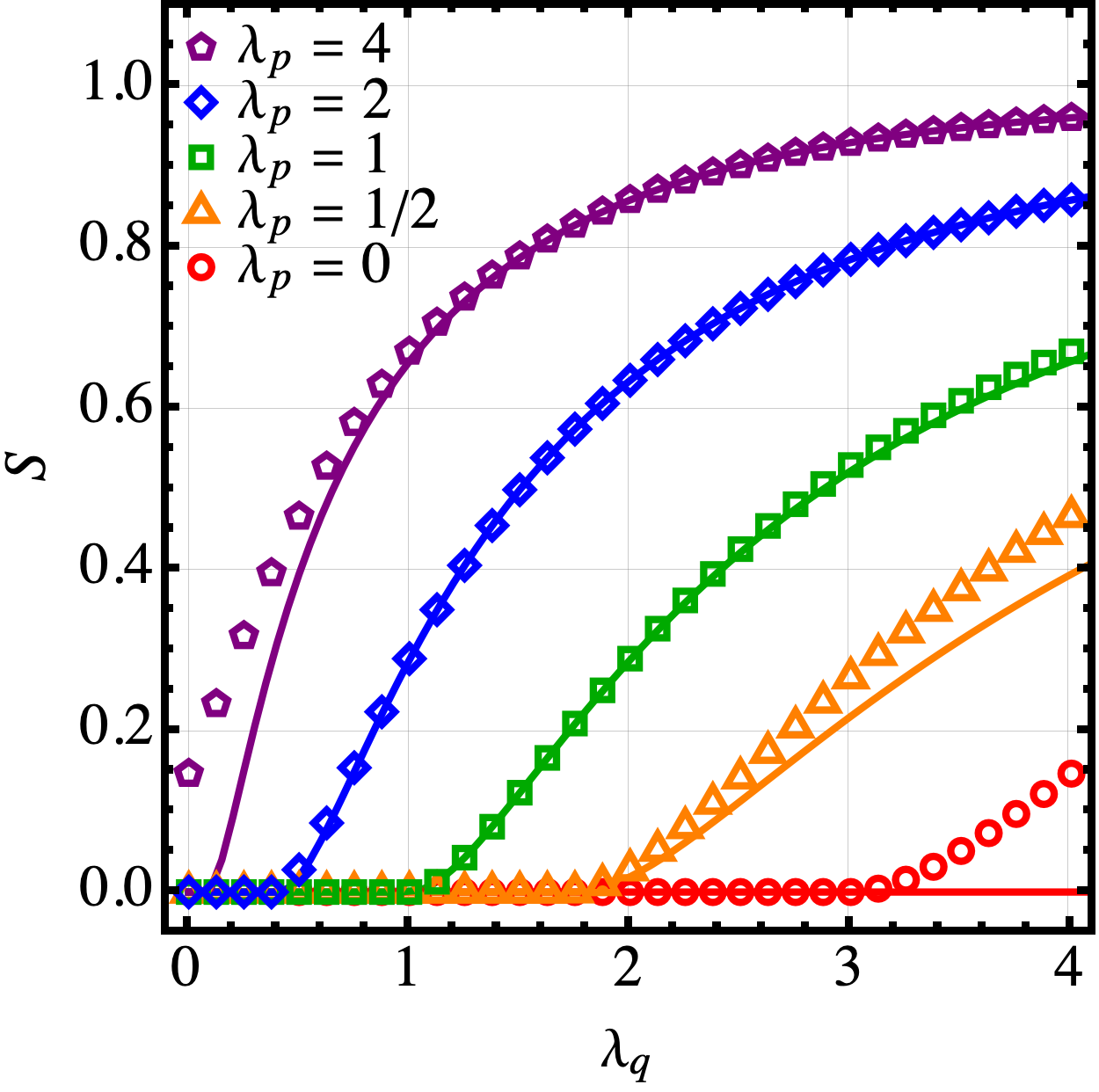}
    
    \caption{Comparison of the value of $S$ predicted by the generating function method (symbols) vs that given by Eq.~(\ref{eq:S transitive}) (solid lines). For values of $\lambda_p$ which are neither too large nor too small, the two approaches give similar results for all values of $\lambda_q$. In contrast, when $\lambda_p$ is large, the generating function method is unable to predict the existence of the phase transition that occurs for small $\lambda_q$. Similarly, for small $\lambda_p$ the generating function method predicts the existence of a GSCC, even when there isn't one.}
    \label{fig:naive S comparison}
\end{figure}

Fig.~\ref{fig:naive S comparison} shows a comparison of the predictions of the generating function method with that given by Eq.~(\ref{eq:S transitive}) in the main body of the paper. We see that for values of $\lambda_p$ which are neither too large nor too small that the two methods appear to agree for almost all $\lambda_q$. In contrast, for large values of $\lambda_p$ the generating function method predicts that a GSCC will always exist, even for extremely small values of $\lambda_q$. This disagrees with the results developed in the main body of the text and validated by simulations in Sec.~\ref{sec:Percolation Simuations}. Similarly, for small values of $\lambda_p$, the generating function method overestimates the size of the GSCC for sufficiently large $\lambda_q$ and even predicts the existence of a GSCC when $\lambda_p=0$ which is impossible. We thus conclude that the generating function method can provide a crude approximation for percolation in structured networks but nothing more.

\begin{figure}
    \centering
    \includegraphics[width=0.8\linewidth]{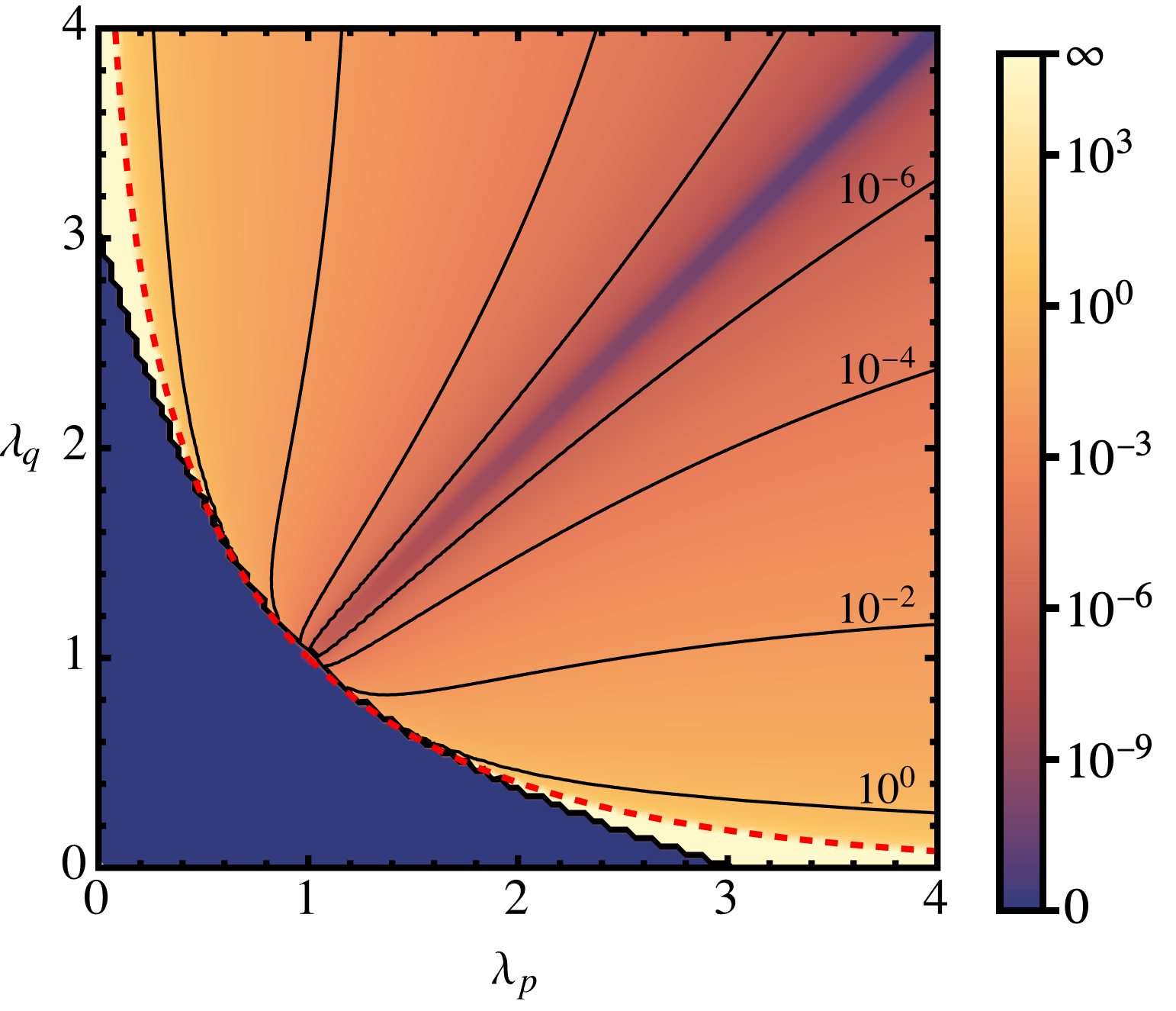}
    
    \caption{The relative error, defined in Eq.~(\ref{eq:rel error}), between $S^\textrm{GF}$, predicted by the generating function method, and $S$, given by Eq.~(\ref{eq:S transitive}). The solid black lines are labelled contours of constant relative error. The dashed red line corresponds to the critical line where the phase transition occurs, as given by Eq.~(\ref{eq:percolation line transitive}). For large values of $\lambda_p$ and small values of $\lambda_q$ or vice versa, the generating function method predicts a percolated phase where there isn't one and thus gives an ``infinite'' relative error. In general, the generating function method does better as we move away from the critical line of phase transition and does extremely well close to the line $\lambda_p=\lambda_q$.}
    \label{fig:rel error}
\end{figure}

More broadly, Fig.~\ref{fig:rel error} shows the relative error 
\begin{equation}
    \textrm{Rel. Error} = 
    \left|\frac{S^\textrm{GF} - S}{S}\right|
    \label{eq:rel error}
\end{equation}
between the value of $S$ predicted by the generating function method, $S^\textrm{GF}$, and the correct value of $S$, given in the main body of the text by Eq.~(\ref{eq:S transitive}). We find that the generating function method is only able to correctly predict the existence of an unpercolated phase over a fraction of the unpercolated region. In particular, as observed in Fig.~\ref{fig:naive S comparison}, large values of $\lambda_p$ with small values of $\lambda_q$ or vice versa cause difficulty for the generating function method. Within the percolated phase, the generating function method tends to do better as we move away from the location of the phase transition and does very well close to the line $\lambda_p=\lambda_q$. This makes sense since near this line, the transitive network becomes extremely similar to the intransitive one, where the generating function method works perfectly.

\clearpage

\bibliography{bib_file}

\end{document}